\documentclass[twocolumn,floatfix,pre,a4paper,showpacs]{revtex4-1}
\pdfoutput=1
\usepackage{graphicx}
\usepackage{amssymb}
\usepackage{amsmath}
\usepackage[utf8]{inputenc}
\usepackage{cancel}
\usepackage[usenames,dvipsnames]{color}
\usepackage{psfrag}
\usepackage{epsfig}
\usepackage{bbm}
\usepackage{bm}
\usepackage{dsfont}
\usepackage{soul}

\newcommand{\gruen}[1]{\textcolor{OliveGreen}{#1}}
\usepackage{colonequals}

\newcommand\abs[1]{\left|#1\right|}

\usepackage{changes}

\usepackage{placeins}

\begin{document}
\title{Chimera patterns under the impact of noise}
\author{Sarah A.~M.~Loos$^1$,  Jens Christian Claussen$^2$, Eckehard Sch\"oll$^1$, Anna Zakharova$^1$}
\affiliation{
  $^1$Institut f\"ur Theoretische Physik,
  Hardenbergstr.~36,
  Technische Universit\"at Berlin,
  D-10623 Berlin,
  Germany
\\
$^2$Computational Systems Biology Lab, Campus Ring 1, Jacobs University Bremen, D-28759 Bremen, Germany
}

\date{\today}
\begin{abstract}
We investigate two types of chimera states, i.e., patterns consisting of coexisting spatially separated domains with coherent and incoherent dynamics, in ring networks of Stuart-Landau oscillators with symmetry-breaking coupling, under the influence of noise. 
Amplitude chimeras are characterized by temporally periodic dynamics throughout the whole network, but spatially incoherent behavior with respect to the amplitudes in a part of the system; they are long-living transients. Chimera death states generalize chimeras to stationary inhomogeneous patterns (oscillation death), which combine spatially coherent and incoherent domains. We analyze the impact of random perturbations, addressing the question of robustness of chimera states in the presence of white noise. 
We further consider the effect of symmetries applied to random initial conditions.
 \end{abstract}
\maketitle

\gruen{%
}

\section{Introduction 
\label{SEC:INTRO}}
%
\textit{Chimera states} are intriguing spatio-temporal patterns made up of spatially separated domains of synchronized (spatially coherent) and desynchronized (spatially incoherent) behavior, although they arise in networks of completely identical units. Kuramoto et al. discovered the chimeras in a network of phase oscillators with a simple symmetric non-local coupling scheme \cite{KUR02a}. Within these ``classical'' chimera states, coherent domains of periodic in-phase oscillations coexist with incoherent domains, characterized by a chaotic behavior in time. In Greek mythology, the chimera is a fabulous creature with a lion's, a goat's, and a snake's head. As the counter intuitive dynamical state is as well composed of incongruous parts, it was named after the beast \cite{ABR04, PAN15}.
Kuramoto's finding initiated a broad wave of theoretical and empirical investigations \cite{ABR08,SET08,LAI09,MOT10,MAR10,OLM10,BOR10,SHE10,SEN10a,WOL11,LAI11,OME11,OME12,OME13,NKO13,HIZ13,SET13,SET14,YEL14,BOE15,BUS15,OME15,OME15a, ASH14}. The first experimental evidence on chimera states was presented only one decade after their theoretical discovery \cite{HAG12,TIN12,MAR13,LAR13,KAP14,WIC13,WIC14,SCH14a,GAM14,ROS14a,LAR15}. In real-world systems chimera states might play a role, e.g., in the unihemispheric sleep of birds and
dolphins~\cite{RAT00}, in epileptic seizures~\cite{ROT14}, in power grids~\cite{MOT13a}, or in social systems~\cite{GON14}. 

The basin of attraction of the chimera states is typically relatively small compared to that of the synchronized state. This explains why they were not detected for a long time. Further, it is the reason why for many investigations specially prepared initial conditions are used. An interesting feature of the chimera states is that they are often long living transients towards the in-phase synchronized oscillatory state.
By this, the coupling between the oscillators introduces a time scale much larger than the 
oscillation periods of each single oscillator.
It was theoretically predicted \cite{WOL11} and experimentally confirmed \cite{ROS14a} that the lifetime of the chimera states of phase oscillators grows exponentially with the system size. This illustrates impressively that they are not simply a temporary trace of the initial conditions, but are a persisting phenomenon.
 Recently, the investigation of chimera states has been generalized towards networks of elements which have more complicated local dynamics \cite{PAN15,BOE15,OME15a}. In particular, nodes which not only involve phase but also amplitude dynamics are considered. As described in \cite{SET13,SET14}, in such systems amplitude-mediated chimeras can be found, which show a chimera behavior with respect to the phases as well as with respect to the amplitudes. In this work, we will consider a different type of chimera states. They are characterized by strictly correlated phase dynamics throughout the whole network, but coexisting domains of coherent and incoherent amplitude dynamics. These \textit{amplitude chimeras} were first described in \cite{ZAK14}. A crucial difference to classical phase chimeras is that the spatial incoherence does not imply chaotic behavior in time. In fact, all nodes of an amplitude chimera perform periodic oscillations, but in the incoherent domain the spatial sequence of the positions of the centers of oscillation is completely random. It has been noted \cite{ZAK15b} that amplitude chimeras are long living transients, however, the dependence of their lifetimes upon the system size is not known. This will be one of the questions addressed in this work.\\
  It has been shown that the occurrence of the amplitude chimeras in a network of Stuart-Landau oscillator is restricted to the case of non-local coupling \cite{ZAK14,ZAK15b}. Furthermore, the breaking  of the rotational $S^1$ symmetry of the single oscillator is crucial. In this work, this symmetry breaking is realized through a coupling term which only involves the real parts instead of the complex variable $z$, in accordance with \cite{ZAK14}.\\ 
This symmetry-breaking coupling introduces a nontrivial spatially inhomogeneous fixed point (steady state) into the system \cite{ZAK13a}, which is typically associated with a pitchfork bifurcation. As a result, oscillation death patterns occur in addition to the oscillatory states. \textit{Oscillation death} is a phenomenon appearing in oscillator networks, when the coupling induces a splitting of a (trivial) homogeneous fixed point into at least two distinct branches \cite{ZAK13a,SCH15b}. This contrasts with amplitude death, where the quenching of the oscillations is caused by stabilization of an unstable spatially homogeneous fixed point which is already present in the uncoupled case \cite{KOS13}.
When the newly born inhomogeneous fixed point is stabilized, typically through an inverse Hopf bifurcation, the oscillatory dynamics can be suppressed. In general, however, oscillation death states can coexist with oscillatory solutions. 
Amplitude chimeras and oscillation death are two different phenomena related to symmetry breaking that occur in the Stuart-Landau oscillator network. Interestingly, a novel type of steady state coherence-incoherence patterns combining features of both phenomena, called \textit{chimera death}, has been discovered in this system \cite{ZAK14}. Like amplitude chimeras, these patterns are characterized by coexisting domains of coherent and incoherent behavior in space. They have recently been extended to globally coupled networks \cite{BAN15,PRE15}.\\

An important question is whether transient chimera patterns persist under perturbations.
 The robustness of classical phase chimeras with respect to heterogeneous natural frequencies of the oscillators \cite{LAI09,LAI09a,LAI10} or heterogeneous connections \cite{LAI12} was investigated. It was found that the specific way in which links are removed from the network defines how strongly the chimera lifetimes are affected. Also in the FitzHugh-Nagumo model, the robustness of chimeras with respect to heterogeneous frequencies and topologies was studied \cite{OME15}. Apart from the present work, the robustness of chimera states with respect to external noise has so far not been considered. This problem is especially relevant in the light of experimental realizations where noise is inevitable. It is known that even at a relatively low intensity, noise can significantly influence the behavior of a nonlinear dynamical system. Under certain conditions noise can cause the increase of coherence, e.g., in coherence resonance, and on the contrary it can also induce irregular, chaotic
  dynamics. Stochasticity appears due to the intrinsic fluctuations in a system, or alternatively can be implemented as an external random control force.
  The influence of noise strongly depends on its characteristics such as intensity, for example, which can be treated as 
bifurcation parameters of a stochastic system. Therefore, the investigation of noisy dynamics is on the one hand significant for the understanding of the processes occurring in nature and on the other hand relevant from the point of view of control. 

In this work, we investigate how external noise influences the occurrence of chimera death states and amplitude chimeras, and how it affects their properties and lifetimes. 
For this purpose, we introduce novel order parameters to characterize chimera states, and discuss the impact of initial conditions. In order to explain our numerical findings, we propose that amplitude chimeras represent a saddle-state in the underlying phase space.\\

The paper is organized as follows. In Sect. II we introduce the model. In Sect. III we analyze amplitude chimeras and chimera death without noise and introduce a global measure which can distinguish between chimera patterns and completely coherent dynamics like in-phase synchronization or traveling waves and can be used to quantify the lifetime of chimeras. Further, we focus on the role of the initial conditions and the dependence of the chimeras upon coupling parameters and system size. In Sect. IV the influence of noise upon the transient times of chimeras is investigated. Sect. V summarizes the results in terms of the different regimes of the coupling parameters, and Sect. VI draws conclusions.

\section{Model}
\label{model}
%
We consider a network of $N$ Stuart-Landau oscillators \cite{KUR02a,ATA03,ZAK13a,ZAK14,ZAK15b} under the impact of external white noise $\xi_j(t)$. The local deterministic dynamics of each node $j\in \{1,...,N \}$ is given by $\dot{z}_j = f(z_j)$, with the normal form of a supercritical Hopf bifurcation
\begin{equation}\label{EQ:SL}
f(z_j) = (\lambda + i \omega -{|z_j| }^2 ) z_j , 
\end{equation}
where $z_j = x_j+i\, y_j \!=\! r_j e ^{i\phi_j} \in \mathbb{C}$, with $x_j,y_j,r_j,\phi_j \in \mathbb{R}$, and $\lambda, \omega > 0$. At $\lambda \!=\! 0$ a Hopf bifurcation occurs, so that for $\lambda > 0$ the single Stuart-Landau oscillator exhibits self-sustained oscillations with frequency $\omega$ and radius $r_j\!=\!\sqrt{ \lambda}$, and the unique fixed point $x_j\!=\!0$, $y_j\!=\!0$ is unstable.
\\
We investigate a ring of $N$ non-locally coupled Stuart-Landau oscillators, where each node is coupled to its $P$ nearest neighbors in both directions with the strength $\sigma > 0$, and is subject to noise of intensity $D > 0$ :
\begin{equation}\label{EQ:SL_network_det}
\dot{z_j} =f(z_j) + \frac{\sigma}{2P} \sum_{k=j-P}^{j+P} (Re z_k - Re z_j) + \sqrt{2D} \xi_j(t) 
\end{equation}
where $j = 1,2,\dots,N$ and all indices are modulo $N$. The coupling and the noise are only applied to the real parts.
In many real-world oscillator systems only one variable is accessible for coupling. For instance, the diffusive coupling through only one variable naturally arises in biological and biochemical systems. The simplest coupling scheme is hence to choose the 
real parts of the complex variable $z$. Here $\xi_j(t) \in \mathbb{R}$ is additive Gaussian white noise \cite{RIS96}, i.e., $\langle \xi_j (t) \rangle \!=\! 0, ~\forall j$, and $\langle \xi_i (t) \xi_j(t') \rangle \!=\!  \delta_{ij} \delta(t-t'), ~\forall i,j$, where $\delta_{ij}$ denotes the Kronecker-Delta and $\delta(t-t')$ denotes the Delta-distribution. Hence the noise is spatially uncorrelated.
\section{Deterministic chimera patterns and the role of initial conditions}
\label{IC}
%
As shown in \cite{ZAK14} in the absence of noise, various different states can be found in the network given by Eq.\,(\ref{EQ:SL_network_det}). Which particular state actually arises, depends on the specific values of the coupling parameters and the initial conditions, as Eq.\,(\ref{EQ:SL_network_det}) describes a multistable system. Among the possible states, two different types of asymptotically stable states can be found, on the one hand oscillatory states, and on the other hand steady state patterns which are related to oscillation death. The latter are represented by completely coherent or completely incoherent oscillation death patterns, as well as by chimera death patterns. The asymptotically stable oscillatory states appear in two different spatio-temporal patterns: in-phase synchronized oscillations and coherent traveling waves. Besides these, long lasting oscillatory transients with interesting features occur, i.e., amplitude chimera states. In this work we demonstrate that
  all these states can also be observed under the influence of noise.
 
%
%
\subsection{Amplitude chimeras and chimera death configurations} \label{SEC:states}
\begin{figure}
\centering
\includegraphics[width=\linewidth]{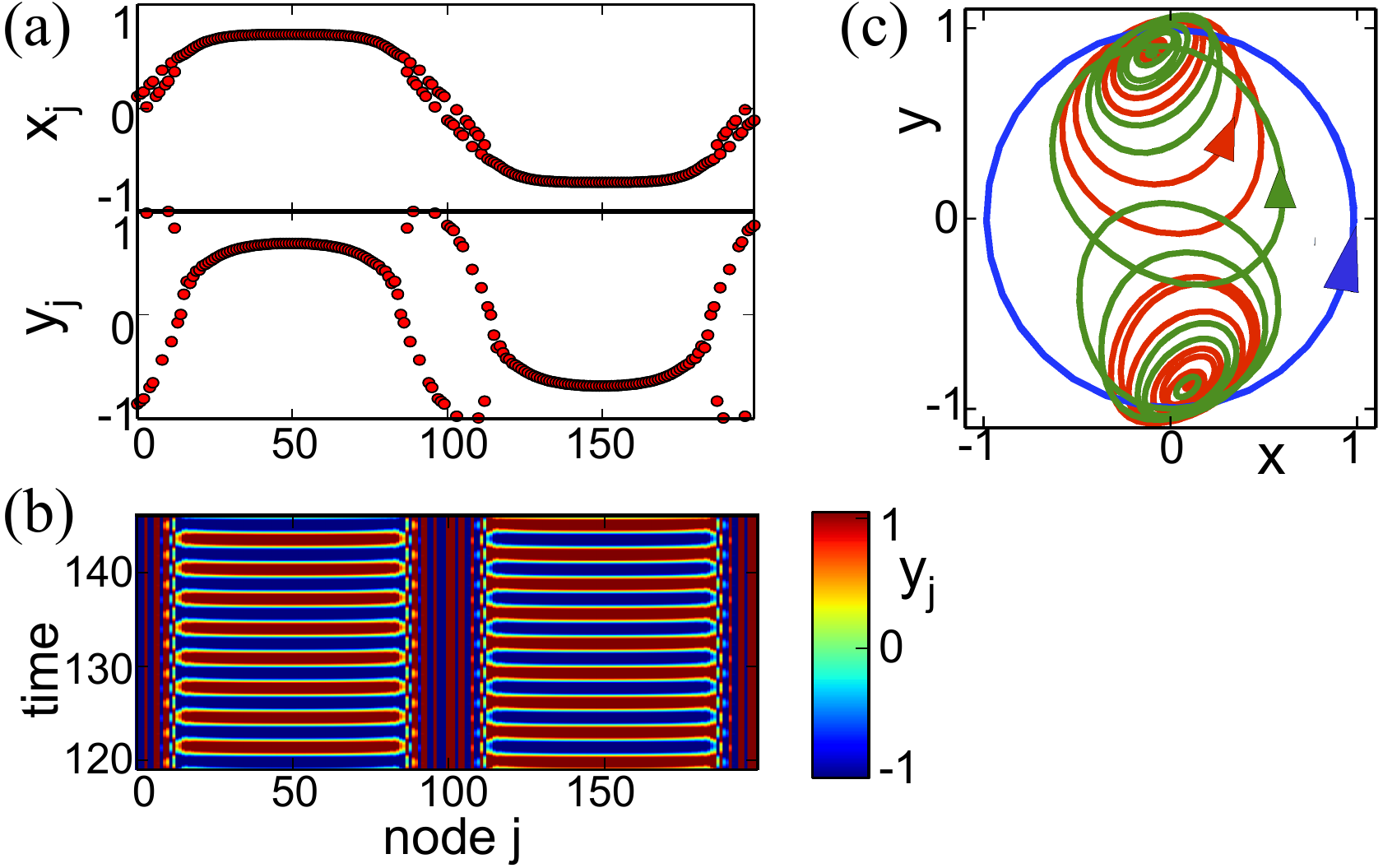}
\caption{(Color online) Amplitude chimera state in system (\ref{EQ:SL}) with $N\!=\!200$ nodes, for coupling range $P/N\!=\!0.04$ and coupling strength $\sigma \!=\!18$: (a) snapshot (top: $x_j$, bottom: $y_j$), (b) space-time plot, (c) phase plot in the complex plane: trajectories of 12 nodes of the coherent domains (unit circle, blue) and 12 nodes of the incoherent domains (red and green), the arrows indicate the direction of the motion. Initial condition: see section \ref{SEC:IC}. Other parameters: $D\!=\!0$, $\lambda\!=\!1$, $\omega\!=\!2$.}
\label{FIG:AC-example} 
\end{figure}
Before an asymptotic oscillatory state (a completely in-phase synchronized oscillation or a coherent travelling wave)
is approached, \textit{amplitude chimera} states can appear as long transients, potentially lasting for hundreds or even thousands of oscillation periods. 
In contrast to classical chimeras, during the chimera all nodes (including the ones within the incoherent domains) oscillate with the same approximate period, $T=\frac{2\pi}{\omega}$, and a spatially correlated phase, but they show spatially incoherent behavior with respect to the \textit{amplitudes} in part of the system. Figure \,\ref{FIG:AC-example} shows an exemplary amplitude chimera configuration. The nodes within the two coherent domains (here $13 \le j \le 85$ and $113 \le j \le 185$) perform synchronized oscillations, all with the same amplitudes. The coherent domains always appear pairwise, such that for every time $t$, all nodes within one coherent domain have opposite phases than all nodes of the other, antipodal domain. Hence they always fulfill the ``anti-phase partner'' condition $z_j = -z_{j+N/2}$, $j\mod N$, assuming even $N$. As visible in Fig.\,\ref{FIG:AC-example} (c), the trajectories in the complex plane of all nodes are cycles, illustrating that all nodes have periodic dynamics in time. This is a fundamental difference between the classical chimera states where a part of the network demonstrates chaotic temporal behavior. The nodes of the coherent domains all oscillate on a perfect circle around the origin. Both coherent domains are represented by one single blue line in Fig.\,\ref{FIG:AC-example} (c), which is at the same time the trajectory of all nodes when the completely in-phase synchronized oscillatory solution is approached. The two antipodal coherent domains are separated by incoherent domains. There, neighboring nodes can be in completely different states at a given time $t$. Their trajectories are deformed circles, whose centers are shifted from the origin. 
In the incoherent domain the sequence of nodes that oscillate around centers in the upper and lower half-plane, respectively, is completely random. The ratio of the domain sizes can vary strongly depending upon the parameters as discussed below, but does not depend upon the initial conditions. Transient amplitude chimeras with very narrow incoherent domains can be observed, as well as with broad ones. \\
\begin{figure}
\centering
\includegraphics[width=\linewidth]{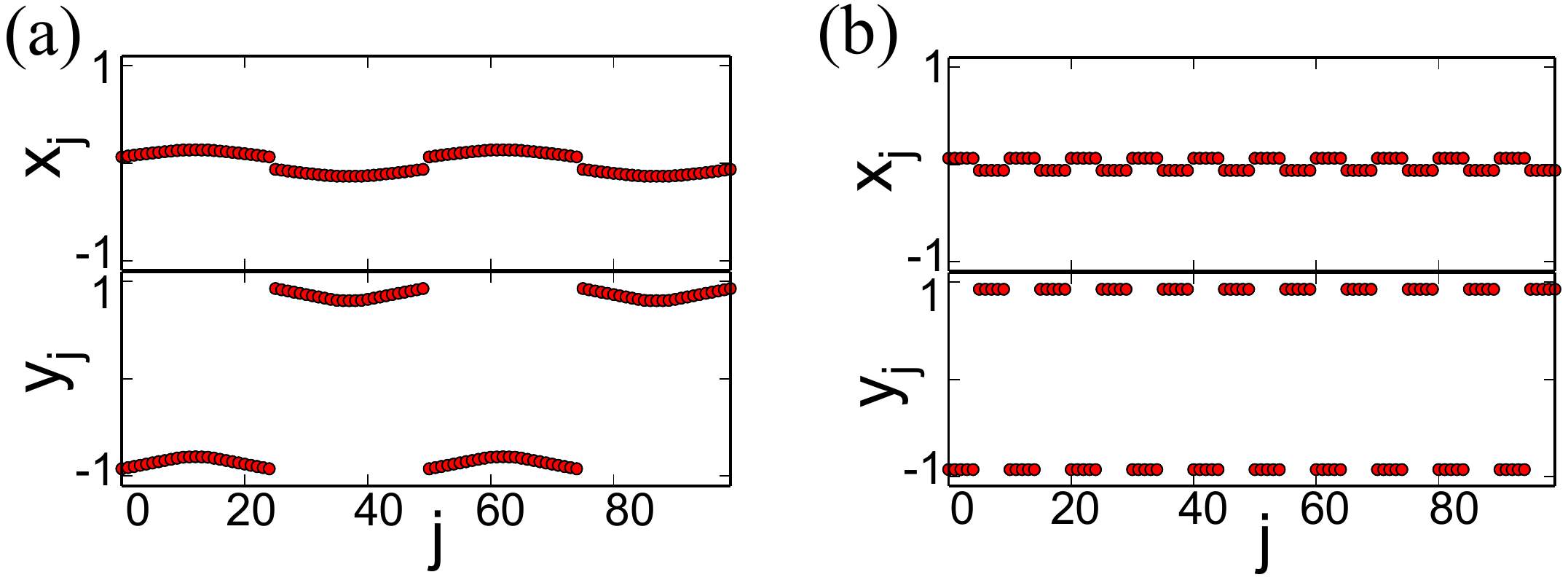}
\caption{(Color online) Snapshots of coherent oscillation death states: (a) coupling strength $\sigma\!=\!18$ and coupling range $P/N=0.14$ (2-cluster), (b) $\sigma\!=\!8$, $P/N\!=\!0.04$ (10-cluster). Initial condition: nodes $0 \le j \le 24$ and $50 \le j \le 74$ are set to $(x_j,y_j)\!=\!(0.1,-1)$, all other nodes are set to $(x_j,y_j)\!=\!(-0.1,+1)$. Other parameters: $N\!=\!100$, $D\!=\!0$, $\lambda\!=\!1$, $\omega\!=\!2$.} 
\label{FIG:OD-examples}
\end{figure}
\begin{figure}
\centering
\includegraphics[width=\linewidth]{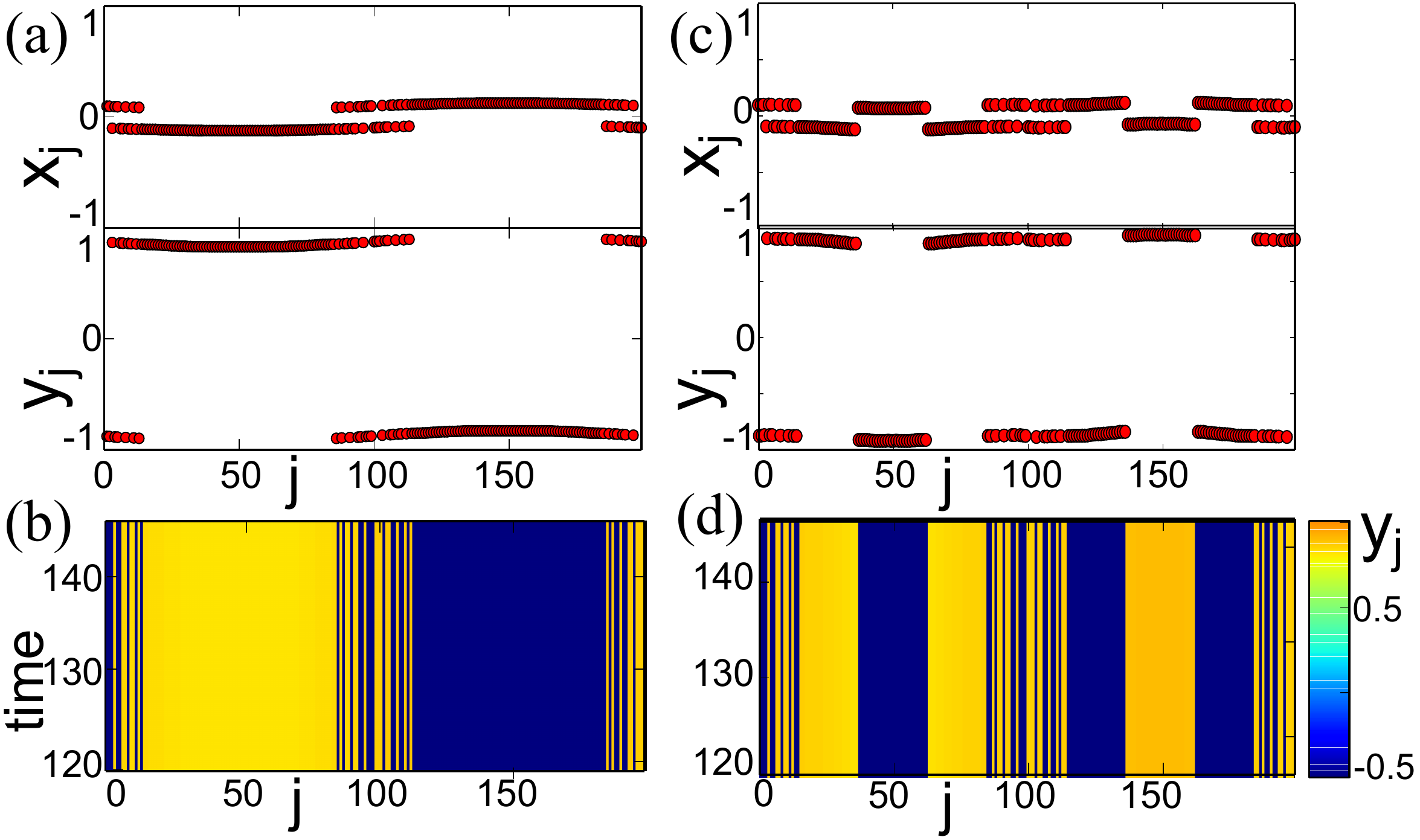}
\caption{(Color online) (a),(b) 1-cluster chimera death (1-CD) for coupling range $P/N\!=\!0.4$, (c),(d) 3-cluster chimera death (3-CD) for $P/N\!=\!0.2$. Snapshots $x_j, y_j$ are shown in panels (a),(c), space-time plots in panels (b),(d). Initial condition: see section \ref{SEC:IC}. Parameters: $N\!=\!200$, $\sigma\!=\!18$, $D\!=\!0$, $\lambda\!=\!1$, $\omega\!=\!2$.}
\label{FIG:1-CD_3-CD}
\end{figure}
If the coupling strength and coupling range exceeds certain values, the oscillations of the Stuart-Landau nodes can be suppressed due to the stabilization of a new inhomogeneous steady state created by the coupling. Instead of performing oscillations, each node approaches a fixed point close to one of the following two branches: $(x^{*1},y^{*1})\!\approx\!(-0.1,+0.85)$ or $(x^{*2}, y^{*2}) \! \approx\!(+0.1,-0.85)$ (for $\lambda=1$), and remains there for all times. The oscillation death states exhibit a huge variety of spatial patterns, including multiple coherent and multiple incoherent oscillation death states \cite{ZAK14,ZAK15b}. Two exemplary configurations of completely coherent oscillation death patterns are shown in Fig.\,\ref{FIG:OD-examples} (2-cluster and 10-cluster oscillation death), see also \cite{SCH15b}. The oscillation death regime is characterized by very high multistability. Among the oscillation death states, \textit{chimera death} patterns can be found, which combine the characteristics
  of both phenomena: chimera state and oscillation death. 
These patterns consist of coexisting domains of coherent and incoherent populations of the inhomogeneous steady state branches. Within the incoherent domains, the population of the two branches (upper and lower) follows a random sequence, as for example visible in Fig.\,\ref{FIG:1-CD_3-CD}. Within the coherent domains, the number of clusters of neighboring nodes that populate the same branch of the inhomogeneous steady state can vary. An $m$-cluster chimera death state ($m$-CD), with $m \in \{1,3,5,7,9,... \}$, is characterized by the occurrence of $m$ clusters within each coherent domain. The coherent domains always appear pairwise with anti-phase symmetry 
$z_j = -z_{j+N/2}$, similarly to the coherent domains of the amplitude chimera configurations. Our numerical results confirm that the stable oscillation death patterns fulfill the ``anti-phase partner'' condition. 
%
\subsection{Order parameter}
\begin{figure*}
	\centering
\includegraphics[width=0.5\linewidth]{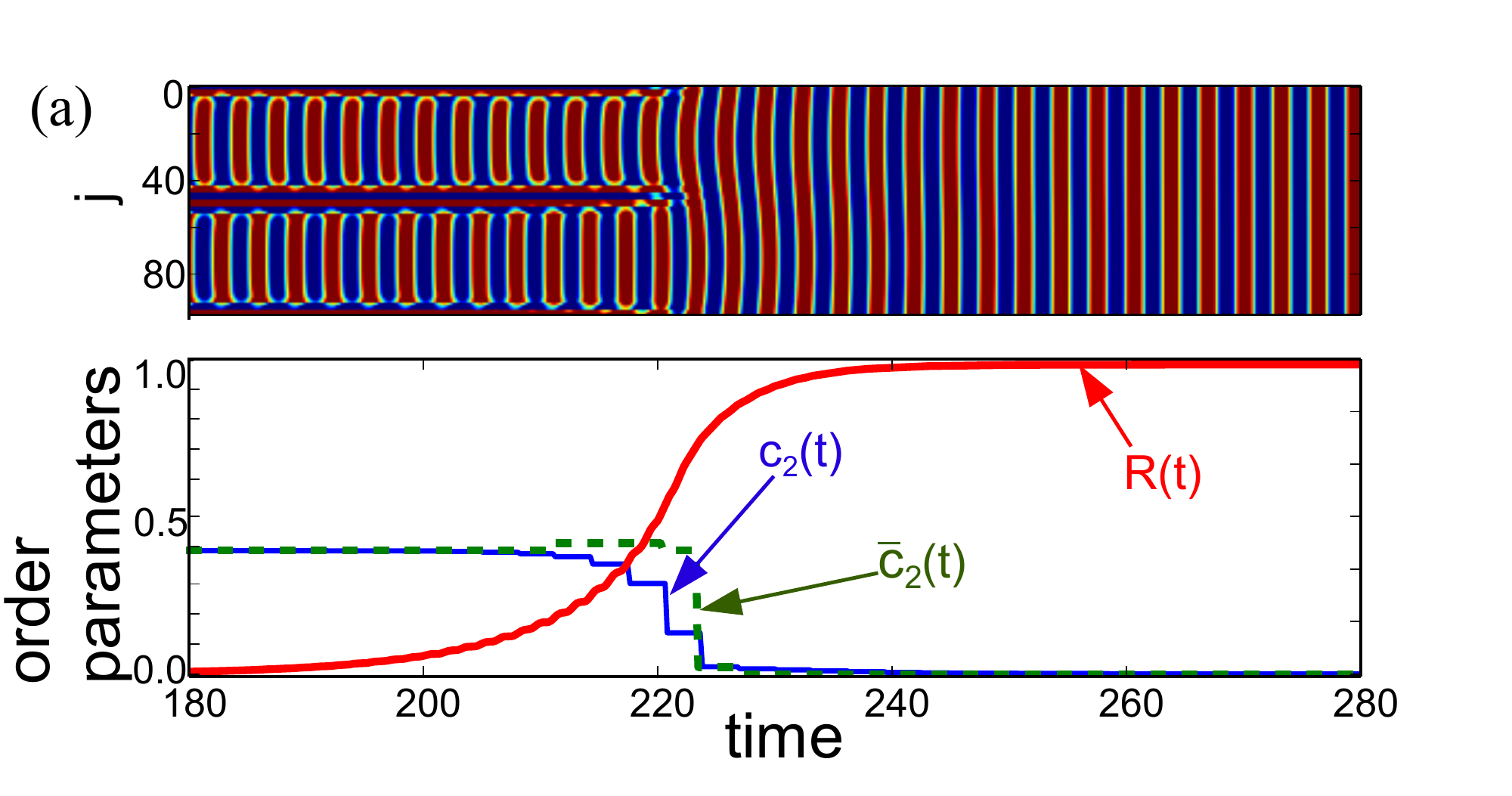}%
\includegraphics[width=0.5\linewidth]{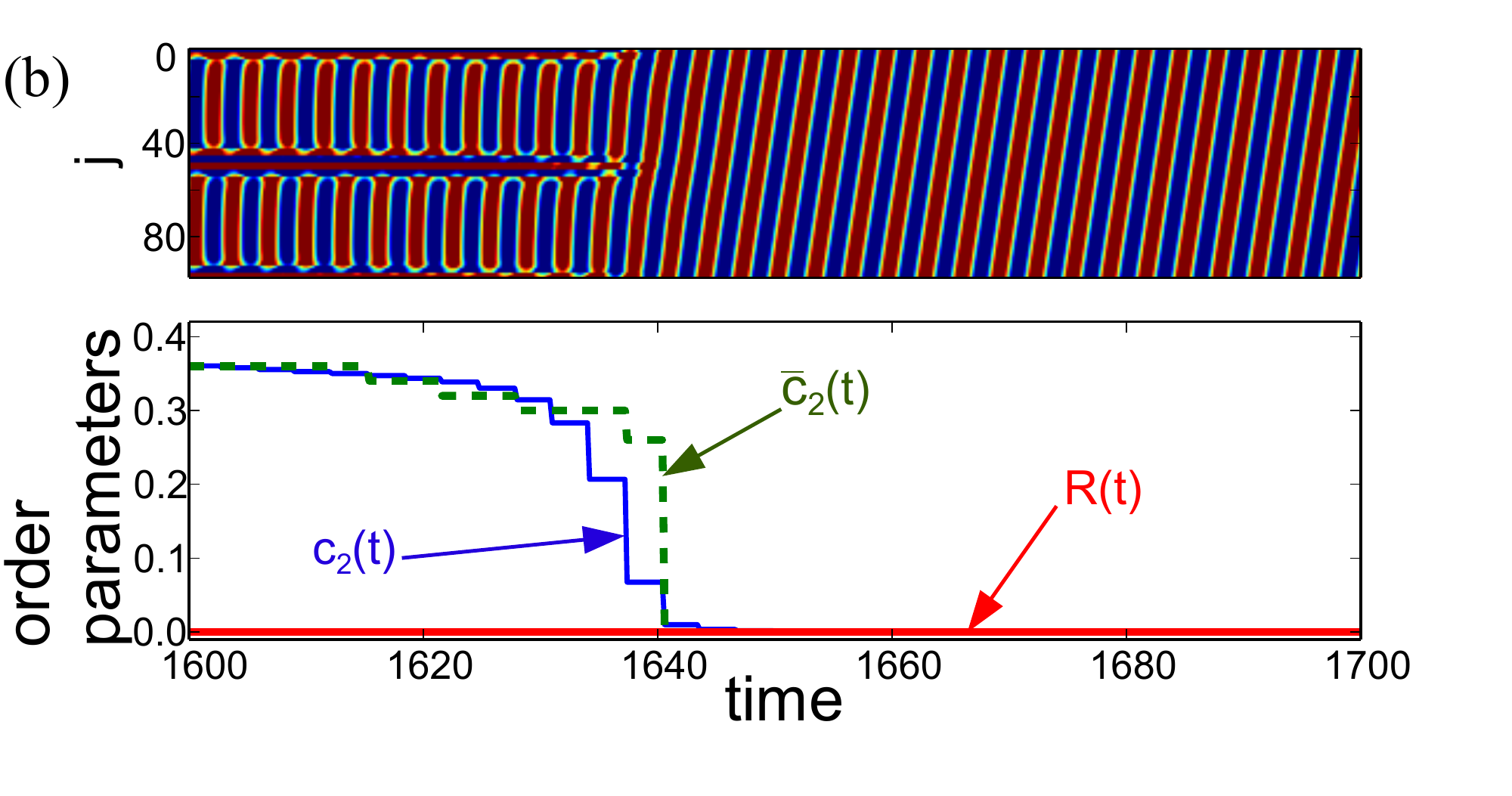}
\caption{(Color online) Transition from an amplitude chimera state to (a) in-phase synchronized oscillation, (b) coherent traveling wave. Upper panels: space-time plots, lower panels: time series of the order parameters: global Kuramoto order parameter $R(t)$, second moments $c_2$, $\overline{c}_2$ with $\alpha\!=\!0.0361$. Parameters $N\!=\!100$, $P/ N\!=\!0.04$, $\sigma\!=\!14$, $D\!=\!0$, $\lambda\!=\!1$, $\omega\!=\!2$. In (a) and (b) different realizations of the initial conditions shown in Fig. 5(b) with $Q=N/4$ are used.\\
} 
\label{FIG:transition_AC_SYNC}
\label{FIG:transition_AC_TW}
\end{figure*}
The global Kuramoto order parameter \cite{KUR02a, ACE05}
\begin{equation} \label{EQ:Def-R(t)}
R(t) = \frac{1}{N} \abs{\sum_{i=1}^{N} z_j(t)} 
\end{equation}
can be used to detect if the system is in the in-phase synchronized oscillatory state, where $R(t)=1$. It is smaller for all other configurations. However, this order parameter cannot be usefully applied to determine the transient time of an amplitude chimera. We note that the in-phase synchronized oscillation is not the only asymptotically stable solution the system can approach; traveling waves are also possible, but $R(t)$ cannot distinguish between amplitude chimeras and coherent traveling waves since both give $R(t)\approx 0$.  
\\
In order to define an appropriate order parameter, we introduce the center of mass $(x_j^{c.m.},y_j^{c.m.})$ of each oscillator $j$ \cite{ZAK14}: 
\begin{equation}\label{DEF:cm}
y_j^{c.m.}= \frac{1}{T} \int_{t}^{t+T} y_j (t') ~ \mathrm{d}t', ~~ \text{with}~~ T=\frac{2\pi}{\omega},
\end{equation}
and an analogous definition of $x_j^{c.m.}$. The shift of the center of mass from the origin is given by:
\begin{equation}
r_j^{c.m.}=\sqrt{(y_j^{c.m.})^2+(x_j^{c.m.} )^2}.
\end{equation}\\ 
These measures allow to distinguish between coherent and incoherent domains: $x_j^{c.m.}$, $y_j^{c.m.}$, and hence also $r_j^{c.m.}$, vanish for nodes within the coherent domains of the amplitude chimera. In contrast, all nodes within incoherent domains have finite values. Here, the spatial sequence of positive and negative signs of $y_j^{c.m.}$ is completely random. Therefore, these quantities can serve as local order parameters.\\

We propose the global order parameter based on their second moments
\begin{equation} \label{EQ:Def-c2(t)}
c_2(t) = \sqrt{\frac{1}{N} \sum_{i=1}^{N} \left(y_i^{c.m.}(t)\right)^2}   
\end{equation}
to detect the transition from an amplitude chimera to a coherent oscillating state. Figure \ref{FIG:transition_AC_SYNC} shows two examples with different initial conditions for such a transition, one where the approached stable asymptotic solution is an in-phase synchronized oscillation, and one where it is a coherent traveling wave. Both figures depict a space-time plot (upper panel), and the temporal evolution of the global order parameters $R(t)$ and $c_2(t)$ for a time series around the transition. As is clearly visible, in contrast to the global Kuramoto order parameter, $c_2$ is capable to detect both types of transition. Another advantage of this order parameter is that its value solely depends on the relative size of the incoherent domains of the amplitude chimera, which in turn depends only on the system parameters. In particular, it does not depend on the pattern of disorder within
the incoherent domains, and hence not on the specific realization of the initial condition. This means that the value $c_2$ is a characteristic quantity of a chimera, which distinguishes it from other patterns.
We define the relative number of nodes within the incoherent domains as the total number of nodes within both incoherent domains normalized by the overall number of nodes within the whole network. To calculate the relative number of nodes within the incoherent domains, we propose a modified version:%
\begin{equation}\label{EQ:Def-overline_c2(t)}
\overline{c}_2(t) = \frac{1}{N} \sum_{j=1}^{N} \Theta \big( (y_j^{c.m.}(t))^2 - \alpha \big) ,
 \end{equation}
 with the Heaviside function $\Theta(x)$, and some fixed threshold $\alpha>0$. The Heaviside function is non-zero only for those nodes where $(y_i^{c.m.}(t))^2$ is above this threshold. For the choice $\alpha \!=\! 0.0361$, the predicted number agrees very well with visual determination of the incoherent domain size, which we have checked with numerous amplitude chimera configurations in a wide range of system parameters.\\
%
\subsection{Role of initial conditions}\label{SEC:IC}
\begin{figure}
\begin{minipage}{0.475\linewidth}
\includegraphics[width=\linewidth]{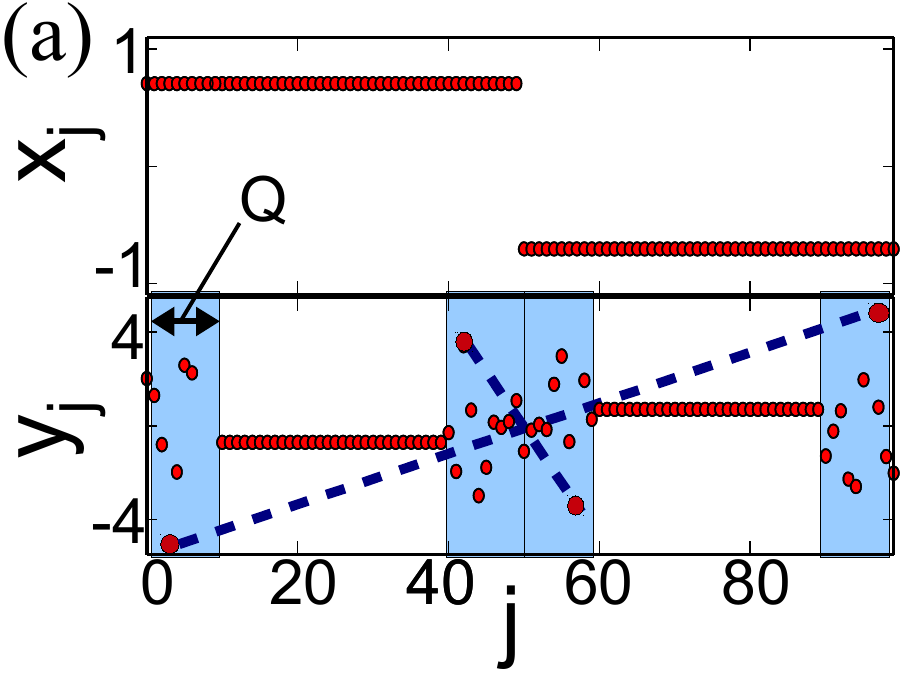}
\end{minipage}
\hspace{0.10cm}
\begin{minipage}{0.475\linewidth}
\includegraphics[width=\linewidth]{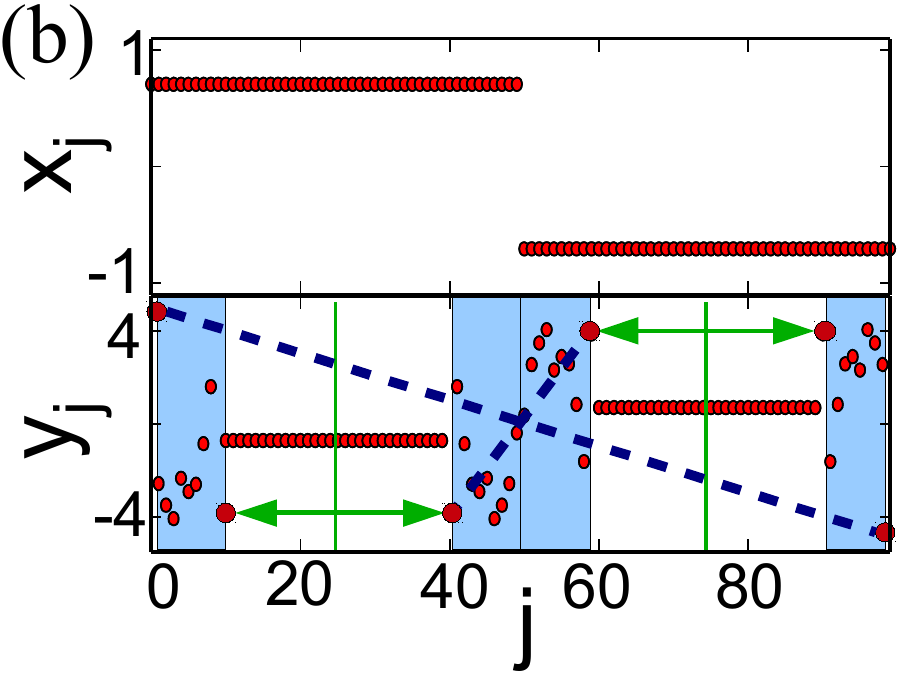}	
\end{minipage}
\caption{(Color online) Specially prepared initial conditions for amplitude chimeras: (a) point-symmetric type, (b) fully-symmetric type. Top panels: $x_j$, bottom panels: $y_j$. Dashed lines indicate the point-symmetry about the center, vertical solid lines and arrows (green) indicate the axial symmetry within both network halves. System size: $N=100$.
} 
\label{FIG:artificialAC_full-symm} 
\label{FIG:artificialAC_point-symm}
\end{figure}
Eq.\,(\ref{EQ:SL_network_det}) is known to describe a multistable system \cite{ZAK14}. Both types of chimera states appear in coupling parameter regimes, where other oscillation death patterns and coherent oscillatory states can be found as well.
   In order to increase the probability of finding chimera states, we use specially prepared initial conditions. 
   A very simple initial condition that produces transient amplitude chimeras in a certain parameter regime (of about $0.01 < P/N < 0.05$, $\sigma<33$), is when all nodes of one half of the network ($1 \le j \le \frac{N}{2}$) are set to the same value $(x_j,y_j)\!=\!(x_0^1,y_0^1)$ (excluding the choice $(0,0)$), and the rest is set to $(x_0^2,y_0^2)=(-x_0^1,-y_0^1)$. Hence, amplitude chimera states can evolve out of an initial configuration that only consists of two completely coherent parts. We choose the values $(x_0^1,y_0^1)\!=\!(\sqrt{0.5}, -\sqrt{0.5})$, so that the nodes start on the limit cycle with $r\!=\!\sqrt{\lambda}=1$, which is the solution for the in-phase synchronized oscillation. The amplitude chimera lifetime nevertheless appears to be of the same order for other values (e.\,g. $(x_0^1,y_0^1)\!=\!(1,-1)$).\\ 

By adding random numbers to $y_j$, we construct a more general class of specially prepared random initial conditions for amplitude chimeras. In particular, we add a random number drawn from a Gaussian distribution with variance $V$ to $y_j$ of the 
$Q$ nodes on the left and on the right side of the borders between both halves (at $j\!=\! \frac{N}{2}$ and $j\!=\!N$), as indicated in Fig.\,\ref{FIG:artificialAC_point-symm} (a), with $Q \in \mathbb{N}$ and $0 < Q \le \frac{N}{4}$. Besides the range $Q$ of incoherence, we also vary  $V\geq 0$. For a proper choice of the two initial condition parameters ($Q$ and $V$), we obtain amplitude chimeras. 
  Using the achieved amplitude chimera lifetime as a quality measure for the initial condition, we compare multiple realizations of the specially prepared random initial conditions for the deterministic system with $P/N\! =\!0.04$ and $N\! =\!100$. We observe that among all considered kinds of initial conditions (different choices for $Q$ and $V$, symmetry conditions, $x_j$ randomized as well, a different underlying distribution for the random numbers), the applied symmetry of the initial condition has the greatest effect upon the transient time.\\

\begin{figure}
\centering
\includegraphics[width=0.8\linewidth]{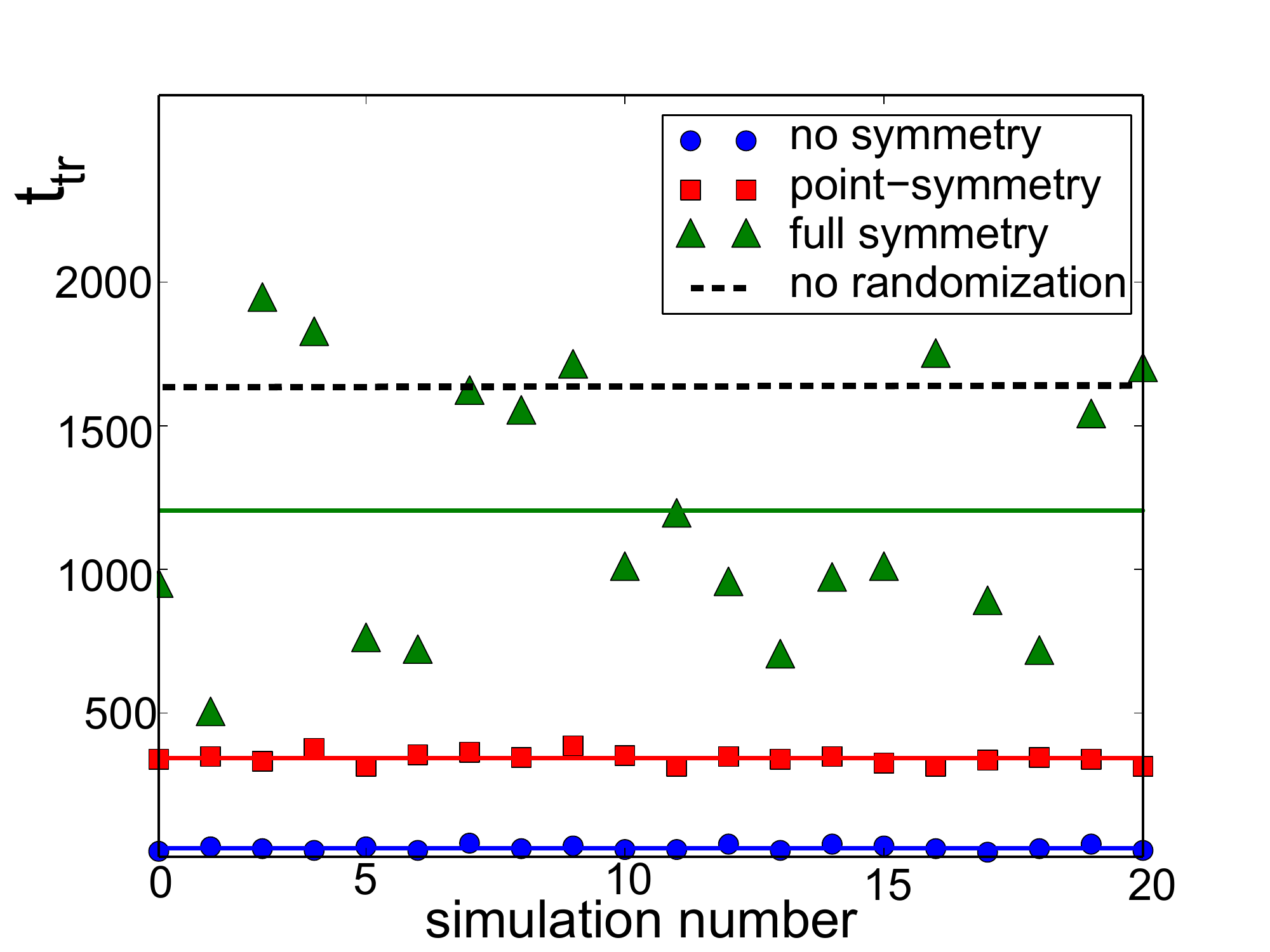}
 \caption{(Color online) Transient times of amplitude chimera states $t_{tr}$ for 20 realizations of specially prepared random initial conditions (no symmetry, point-symmetry, full symmetry, depicted by different symbols) with $Q\!=\!10$, $V\!=\!2$.
 Horizontal solid lines: mean values. Dashed line: $t_{tr}$ for the initial condition with $V\!=\!0$ (no randomization). System parameters: $N\!=\!100$, $\sigma\!=\!14$, $P/N\!=\!0.04$, $D\!=\!0$, $\lambda\!=\!1$, $\omega\!=\!2$.} 
 \label{FIG:IC_compare-symm}
\end{figure}
Figure \ref{FIG:IC_compare-symm} shows the transient times and their mean value (solid lines) for multiple realizations of the initial conditions following three different symmetry schemes. For the particular choice ($Q\!=\!10$, $V\!=\!2$) all symmetry types lead on average to shorter lifetimes than an initial condition without random component (black dashed line). For the initial configurations without symmetry, a random number is chosen independently for each node within the four incoherent intervals. These configurations clearly create the shortest amplitude chimera lifetimes, lasting at most for a couple of oscillation periods. This symmetry type also leads to the shortest transients in other regimes of $Q$, and $V$ (not shown here). In contrast, for the point-symmetric initial conditions, we mirror the random numbers used for $j\in \{ 1,...,\frac{N}{2} \}$ with respect to the center $j\!=\!0$, $y_j\!=\!0$, and use their negative counterparts for the second half. We hence only generate $2 Q$ random numbers in total. The initial configurations are point-symmetric with respect to the center, see Fig.\,\ref{FIG:artificialAC_point-symm} (a). The lifetimes of the occurring amplitude chimeras are much longer than in the non-symmetric case. However, the symmetry type which leads to the longest lifetimes is the one referred to as full symmetry; the initial conditions fulfill two symmetries: The randomly chosen values of the positions of the nodes within the first incoherent interval $1 \le j\le Q$ are mirrored to the nodes $\frac{N}{2}-Q \le j \le \frac{N}{2}$, by setting: $z_j\!=\!z_{\frac{N}{2}+1-j}$. To obtain the positions of the second network half, then a phase shift of $\pi$ is applied, such that the ``anti-phase partner'' condition is fulfilled ($z_j\!=\!-z_{j+\frac{N}{2}}$ and $j\mod N$). Thus, we only generate $Q$ different random numbers in total. The configurations are again point-symmetric with respect to the center, and have an additional axial symmetry with orthogonal axes through $j\!=\!\frac{N}{4}$ and $j\!=\!\frac{3 N}{4}$, as indicated in Fig.\,\ref{FIG:artificialAC_full-symm} (b). Of course, the simple initial condition with no randomization also fulfills these symmetry conditions and can therefore be regarded as one special type of the fully-symmetric specially prepared initial conditions (with $V\!=\!0$).
We have further tested another type of initial condition that solely fulfills the anti-phase partner condition: $z_j\!=\!-z_{j+\frac{N}{2}}$, but has no other symmetries. 
  This type of initial condition also certainly leads to transient amplitude chimeras, but only within very narrow ranges of $Q$ and $V$. For $Q\!=\!10$, $V\!=\!2$, the mean lifetime (of about $t_{tr}\!\approx \!49$), is only slightly increased compared to the non-symmetric initial condition (not shown here). \\
Since the symmetry which is applied to the initial conditions remains preserved during the dynamic evolution, this observation means that the fully-symmetric amplitude chimeras are most stable and have the longest lifetimes. 
\\

\begin{figure}
\includegraphics[width=\linewidth]{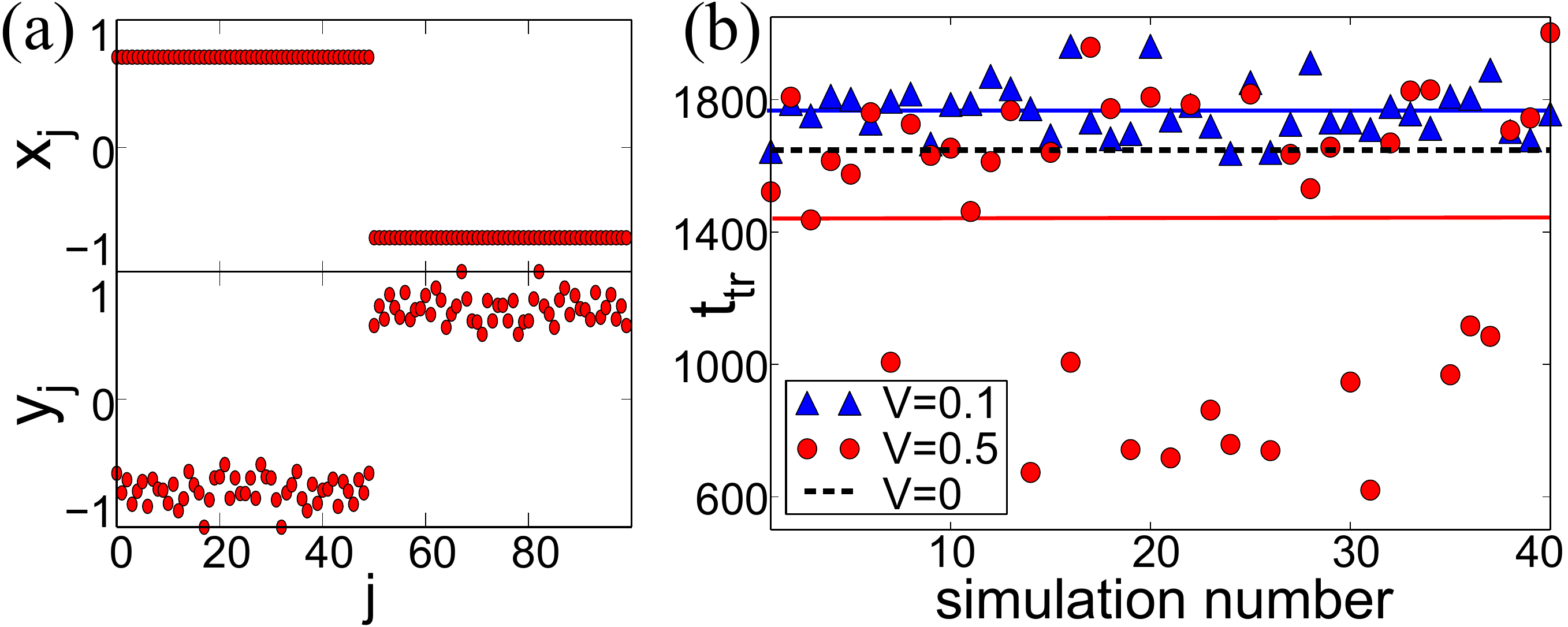}
\caption{(Color online) (a) Fully-symmetric random initial condition with range of incoherence $Q\!=\!\frac{N}{4}$ and variance $V\!=\!0.1$. (b) Transient times of amplitude chimeras $t_{tr}$ for 40 realizations of the initial condition shown in (a): circles: $V\!=\!0.5$, triangles: $V=0.1$. Horizontal solid lines: mean values, dashed line: $t_{tr}$ for $V=0$. System parameters: $N\!=\!100$, $\sigma\!=\!14$, $P/N\!=\!0.04$, $D\!=\!0$, $\lambda\!=\!1$, $\omega\!=\!2$.} 
\label{FIG:artificialAC_R=25_V=0-1}
\label{FIG:IC_compare-V}
\end{figure}

By decreasing the variance in the interval $0.1 \le V \le 2$, the mean amplitude chimera lifetimes increase. In the range of small variances of about $V<0.5$, amplitude chimeras occur for all choices of the incoherence range $Q$, and the particular choice of $Q$ does not influence the transient time much. For $Q\! =\!\frac{N}{4}$, all nodes are randomized, see Fig.\,\ref{FIG:IC_compare-V} (a), which
appears to be a natural choice. 
  Fig.\,\ref{FIG:IC_compare-V}(b) shows the corresponding transient times belonging to a set of 40 realizations of the specially prepared random initial condition with $Q\! =\!\frac{N}{4}$ and $V\!=\!0.5$, and for a set with $V\!=\!0.1$. The mean transient times are much longer than for $V\!=\!2$ (cf. Fig.\,\ref{FIG:IC_compare-symm}). They are at least of the same order (and can be larger) as the transient time for the simple initial condition with no randomization, $V\!=\!0$ (dashed black line). For the choice $V=0.1$, the transient times are increased as compared to $V\!=\!0.5$.
We use the fully-symmetric initial conditions with $Q\! =\!\frac{N}{4}$ and $V\!=\!0.1$ for all investigations presented in this paper. \\
  \\
  Besides oscillatory states, oscillation death states can occur in a large variety of different spatial patterns. Our numerical results suggest that in the appropriate parameter regime every amplitude chimera snapshot can be used as initial condition to certainly produce a chimera death state. How many clusters in the coherent domain of the chimera death pattern occur, depends on the initial condition as well as on the parameter choice (see section \ref{SEC:maps}).
%
\subsection{Relative size of the incoherent domains} \label{SEC:IncDomain}
The ratio of coherent and incoherent domains of an amplitude chimera can be measured by the number of nodes within the incoherent domains divided by the total number of nodes.  It can be calculated using the order parameter $\overline{c}_2(t)$. Interestingly, in the time evolution of the system, this quantity typically fluctuates by less than one order of magnitude, during the chimera, as well as during the coherent synchronized oscillation, or during the traveling wave. This becomes visible by considering the temporal evolution of $\overline{c}_2(t)$, as exemplarily shown in Fig.\,\ref{FIG:transition_AC_SYNC}. In contrast to $R(t)$ and $c_2(t)$, the values of $\overline{c}_2(t)$ show a sharp drop to zero when the amplitude chimera disappears, illustrating that the relative size of the incoherent domains remains constant until the last oscillation period before the transition to complete coherence.\\

Note that the relative size of the incoherent domains is not related to the choice of $Q$, but is completely independent of the initial condition. Even if the incoherent domain is chosen larger or smaller 
initially, it evolves into a domain with a characteristic size corresponding
to the given set of parameters. This is an important general distinguishing factor of chimeras which emphasizes that 
they are self-organized patterns.
\\

\begin{figure}
\centering
\includegraphics[width=0.6\linewidth]{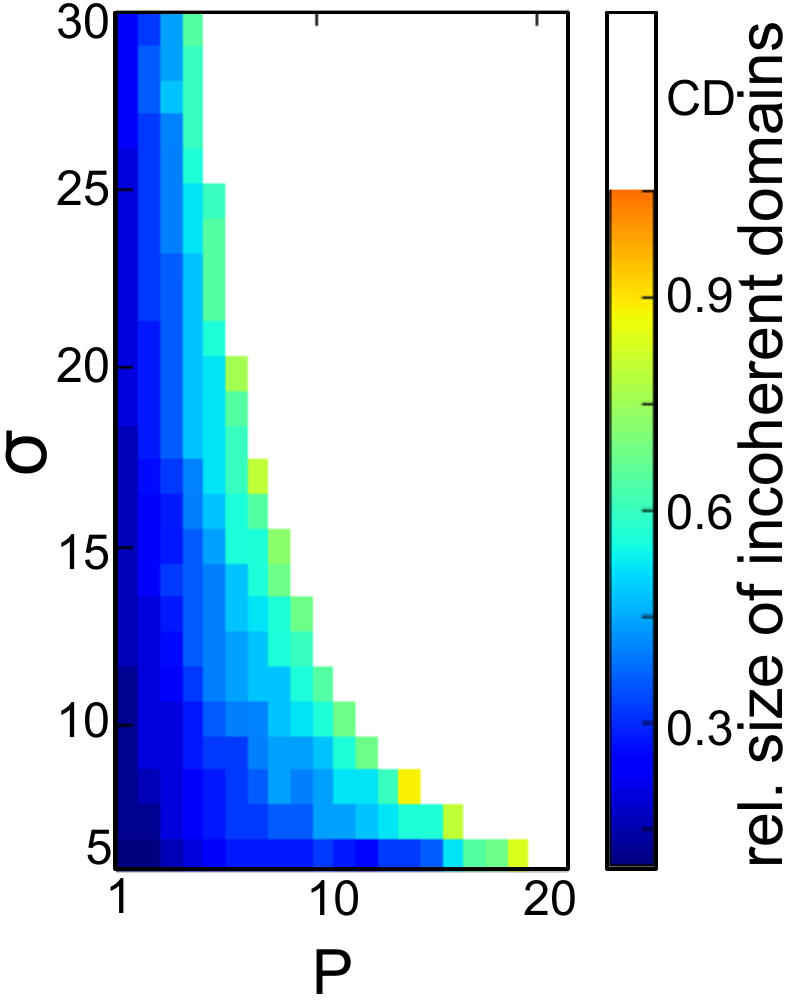}
\caption{(Color online) Relative size of the incoherent domains in the plane of coupling strength $\sigma$ and nearest neighbors number $P$. Initial condition: a fixed snapshot of an amplitude chimera. The light grey area denotes chimera death (CD). Other parameters: $N\!=\!100$, $D\!=\!0$, $\lambda\!=\!1$, $\omega\!=\!2$.}
\label{FIG:incDomain_imshow} 
\end{figure}
\begin{figure}
\centering
\includegraphics[width=0.85\linewidth]{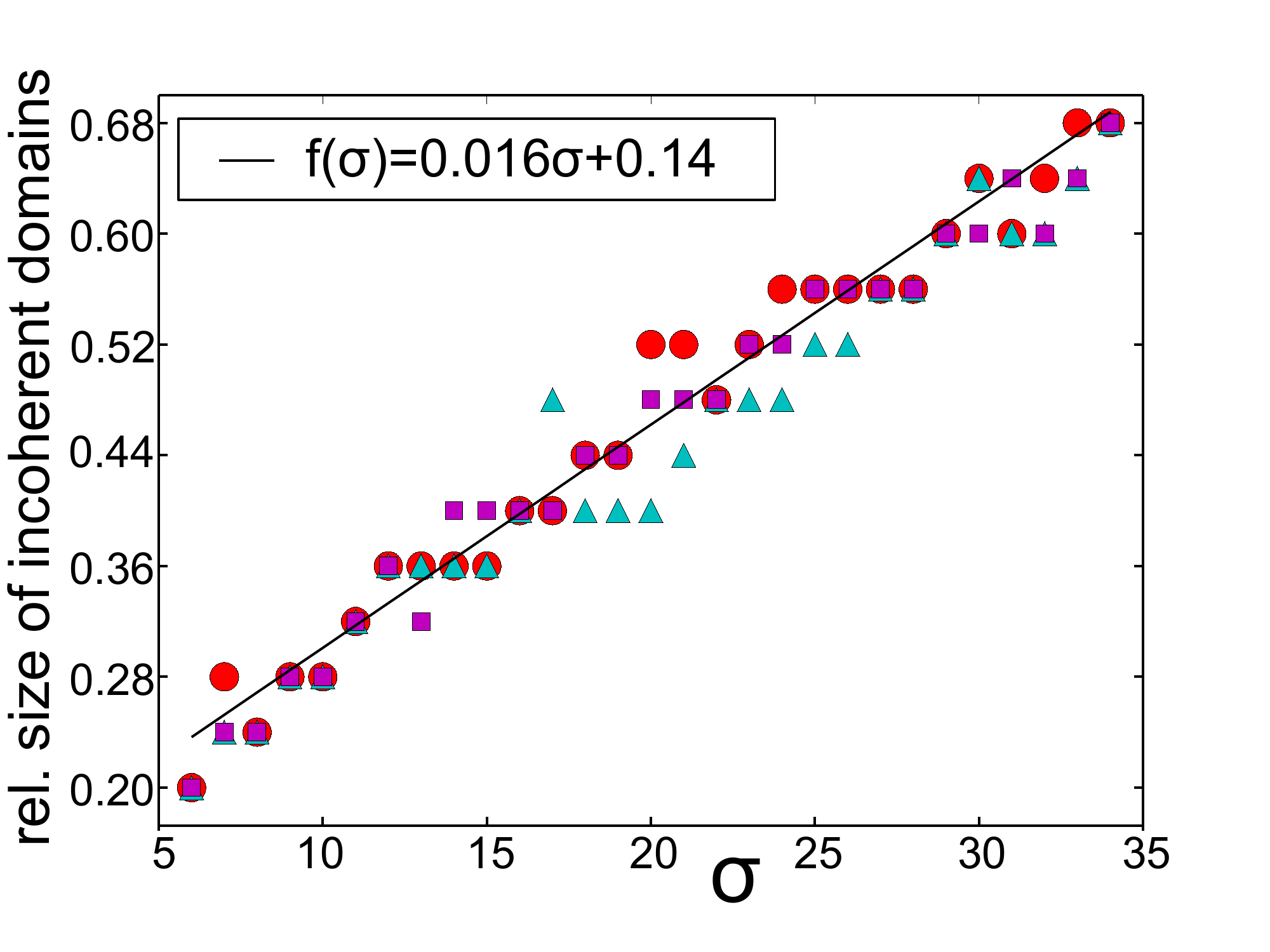}
	\caption{(Color online) Relative size of the incoherent domains of amplitude chimeras vs. coupling strength $\sigma$, measured by $\overline{c}_2(t\!=\!100)$. Squares, triangles, and disks correspond to three different initial conditions. Solid line: linear fit to disks. System parameters: $N\!=\!100$, $P/N\!=\!0.04$, $D\!=\!0$, $\lambda\!=\!1$, $\omega\!=\!2$. 
}
	 \label{FIG:incDomain-sigma}
\end{figure}
\begin{figure}
\begin{center}
\includegraphics[width=\linewidth]{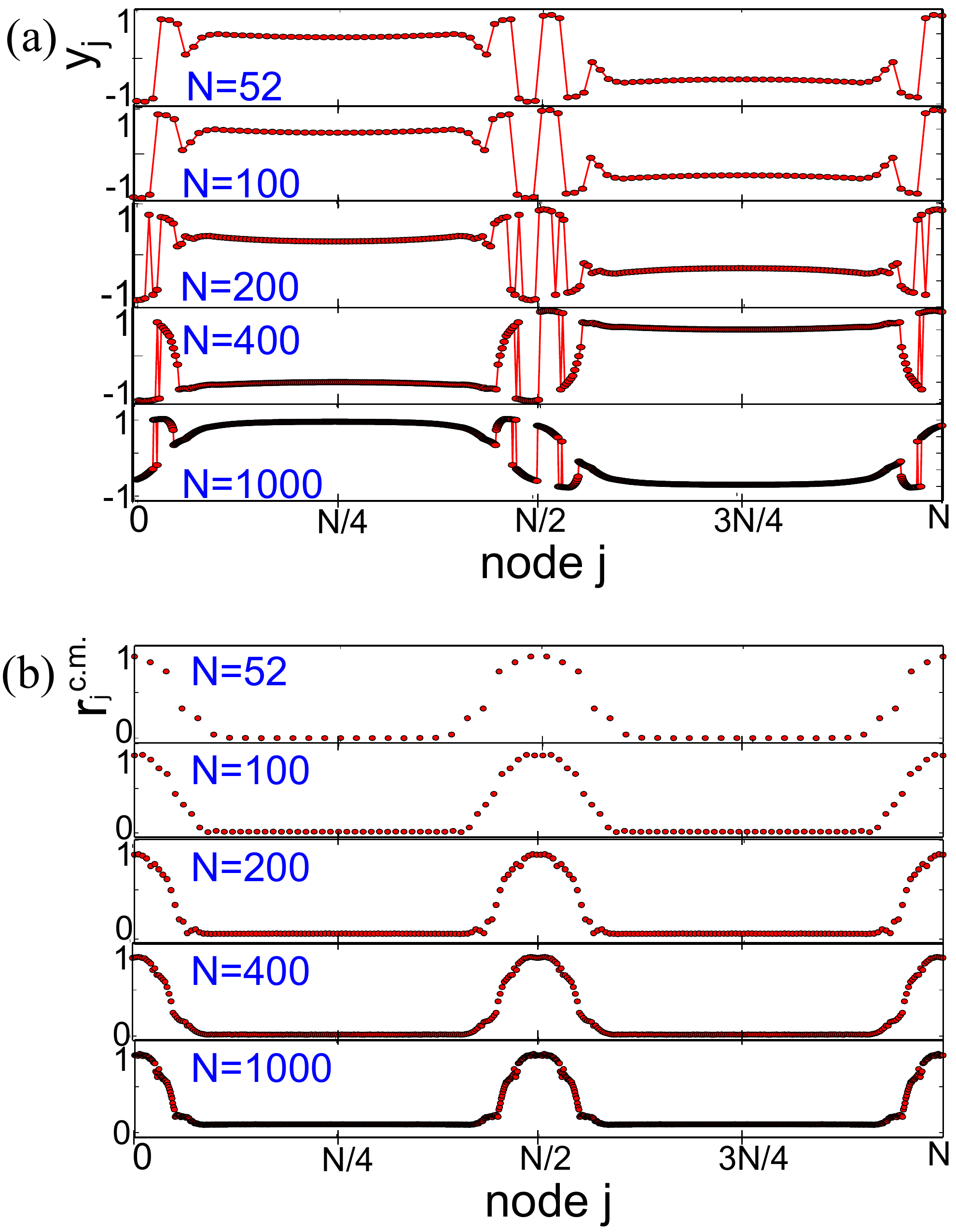}
\caption{(Color online) Snapshots of amplitude chimeras for different system sizes $N$, but constant coupling range $P/N\!=\!0.04$: (a) $y_j$ (b) shift of the center of mass $r_j^{c.m.}$ from the origin. Parameters: $t=140$, $\sigma\!=\!14$, $D\!=\!0$, $\lambda\!=\!1$, $\omega\!=\!2$. Initial conditions: see section \ref{SEC:IC}.
}
\label{FIG:ACsnapshots-variousN}
\label{FIG:ACrcm-variousN}
\end{center}
\end{figure}

In contrast, the relative size of the incoherent domains does depend on the choice of the coupling parameters. This is shown in Fig.\,\ref{FIG:incDomain_imshow} for coupling parameters varied in the range $1\! \le\! P\! \le \! 20$, and $5\! \le\! \sigma\! \le\! 30$. 
 For each choice or the coupling parameters, the same snapshot of an amplitude chimera is used as initial condition. The maximum simulation time is $t=5000$. 
   The relative size of the incoherent domains varies over a wide range from about $0.14$ to about $0.6$, i.e., we observe amplitude chimeras where just $14\%$ of the nodes show a spatially incoherent behavior, up to configurations with $60\%$. At the border between the oscillatory regime and the chimera death regime, we detect the amplitude chimeras with the largest incoherent domains. The width of the incoherent domains appears to increase linearly with $P$, as well as with $\sigma$. In Fig.\,\ref{FIG:incDomain-sigma} the relative size of the incoherent domains in dependence upon the coupling strength is depicted for three different initial conditions. The actual spread of the values is due to the fact that, firstly, $\overline{c}_2(t)$ can assume only multiples of $\frac{1}{N}$, and, secondly, 
 the relative number of nodes can change only in multiples of $\pm \frac{4}{N}$, due to the symmetry in the initial conditions (see section \ref{SEC:IC}).
As opposed to the linear increase with $\sigma$ and $P$, the relative size of the incoherent domains is roughly constant for all system sizes. For an exemplary choice of the coupling parameters, Fig.\,\ref{FIG:ACsnapshots-variousN} shows (a) snapshots and (b) the corresponding shifts from the origin of the centers of mass, for five different system sizes $N$.
%
%
%
\subsection{System size dependence}
For classical phase chimeras it has been established that the transient nature of the chimera states is a finite-size effect \cite{PAN15, WOL11}. In the limit of an infinitely large system, the phase chimera is stable. This is opposite to the system size dependence of the amplitude chimera states, where we find that their lifetime decreases with $N$.\\

\begin{figure}
\centering
\includegraphics[width=0.8\linewidth]{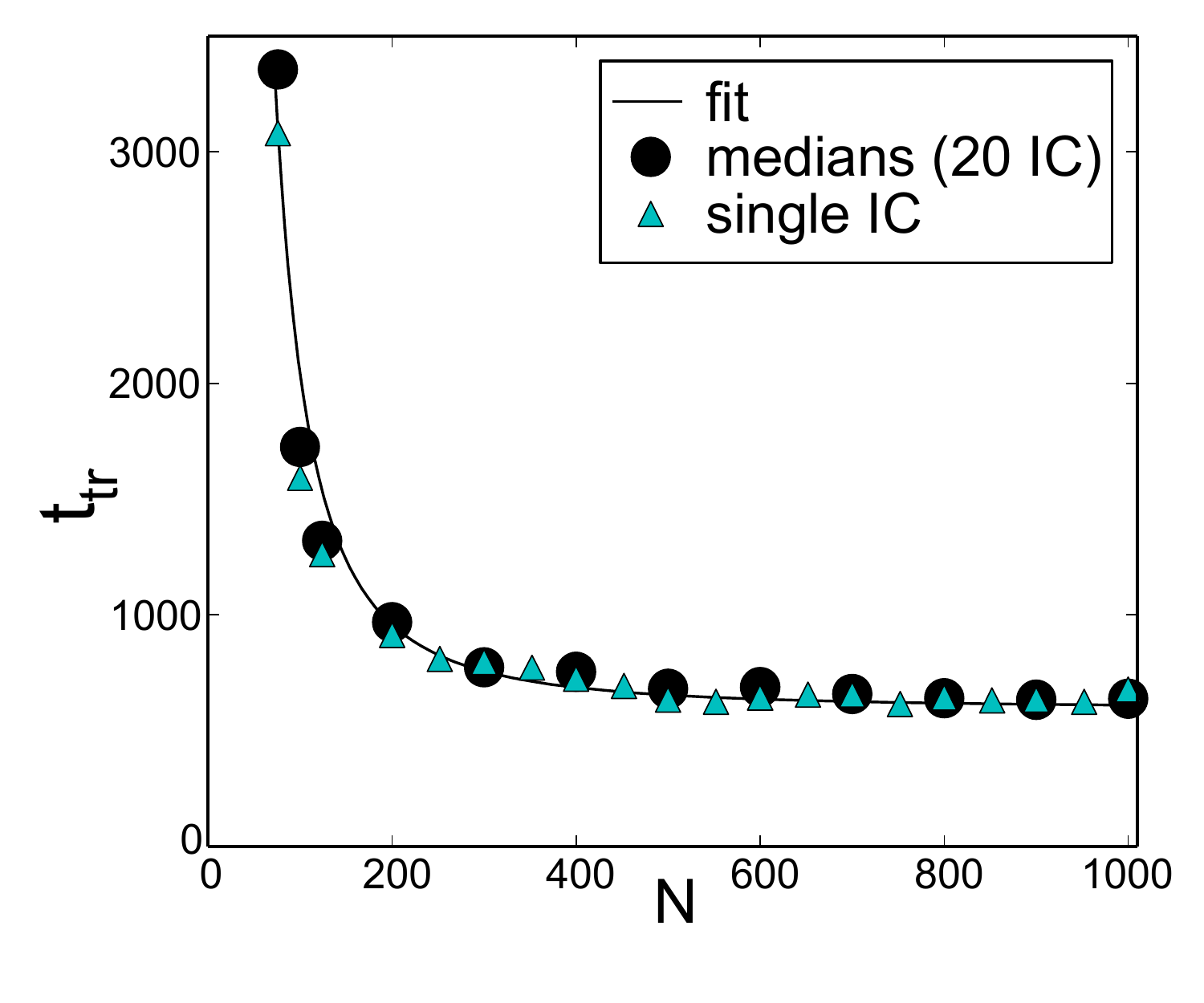}
\caption{(Color online) Transient times of amplitude chimeras $t_{tr}$ vs. system size $N$, for the constant coupling range $P/N=0.04$. Disks: medians of $t_{tr}$ over 20 initial conditions (IC), solid line: fit to Eq.\,(\ref{EQ_transient-N}) with $\alpha_1\approx = 3802.29$, and $\alpha_2\approx 596.23$. Triangles: $t_{tr}$ for a single initial condition. Other parameters: $\sigma\!=\!14$, $D\!=\!0$, $\lambda\!=\!1$, $\omega\!=\!2$.
} 
\label{FIG:mean-transient-times-different-N}
\end{figure}

For various system sizes $N$, Fig. \ref{FIG:mean-transient-times-different-N} shows the median of the transient times for a constant coupling strength $\sigma\!=\!14$ and a constant coupling range $P/N\!=\!0.04$, averaged over 20 initial conditions (black disks).  The data for an additional single initial condition are also shown by cyan triangles. Both data sets can be well described by the following dependence of the amplitude chimera lifetime on the system size:
\begin{equation}\label{EQ_transient-N}
t_{tr}(N)=\left( \frac{\alpha_1}{N}\right)^2 + \alpha_2.
\end{equation}
The black curve in Fig. \ref{FIG:mean-transient-times-different-N} is a fit with $\alpha_1\!\approx\!= 3802.29$ and $ \alpha_2\!\approx\!596.23$. This indicates that amplitude chimeras should exist, albeit with shorter lifetime, in systems of every size. That means that they would also appear in the ``thermodynamic limit'' of infinitely large systems with a finite lifetime (for instance, $t_{tr}\! \approx \! 600$ for $N \rightarrow \infty$, for the type of initial condition used in Fig.~\ref{FIG:mean-transient-times-different-N}). 
These results show that the mechanism of the transient is different from phase chimeras, where the transition is triggered by a sweeping front which travels through the whole system. Rather, the incoherent oscillators of the amplitude chimeras catch up locally with the coherent ones to synchronize, which is independent of the global system size as long as this size is large enough \cite{ZAK15b}.
This reasoning is in line with our observations in Sect. \ref{SEC:IncDomain} that the relative size of the incoherent domains firstly, is roughly constant for all $N$ (Fig.\,\ref{FIG:ACsnapshots-variousN}), and secondly, stays constant during the temporal evolution of a chimera until the last period before the transition to the asymptotic solution and then drops immediately to zero (Fig.\,\ref{FIG:transition_AC_TW}). 
%
%
%
\section{Influence of noise on transient times}
\label{noise}
\begin{figure}[!ht]
\centering
\includegraphics[width=0.8\linewidth]{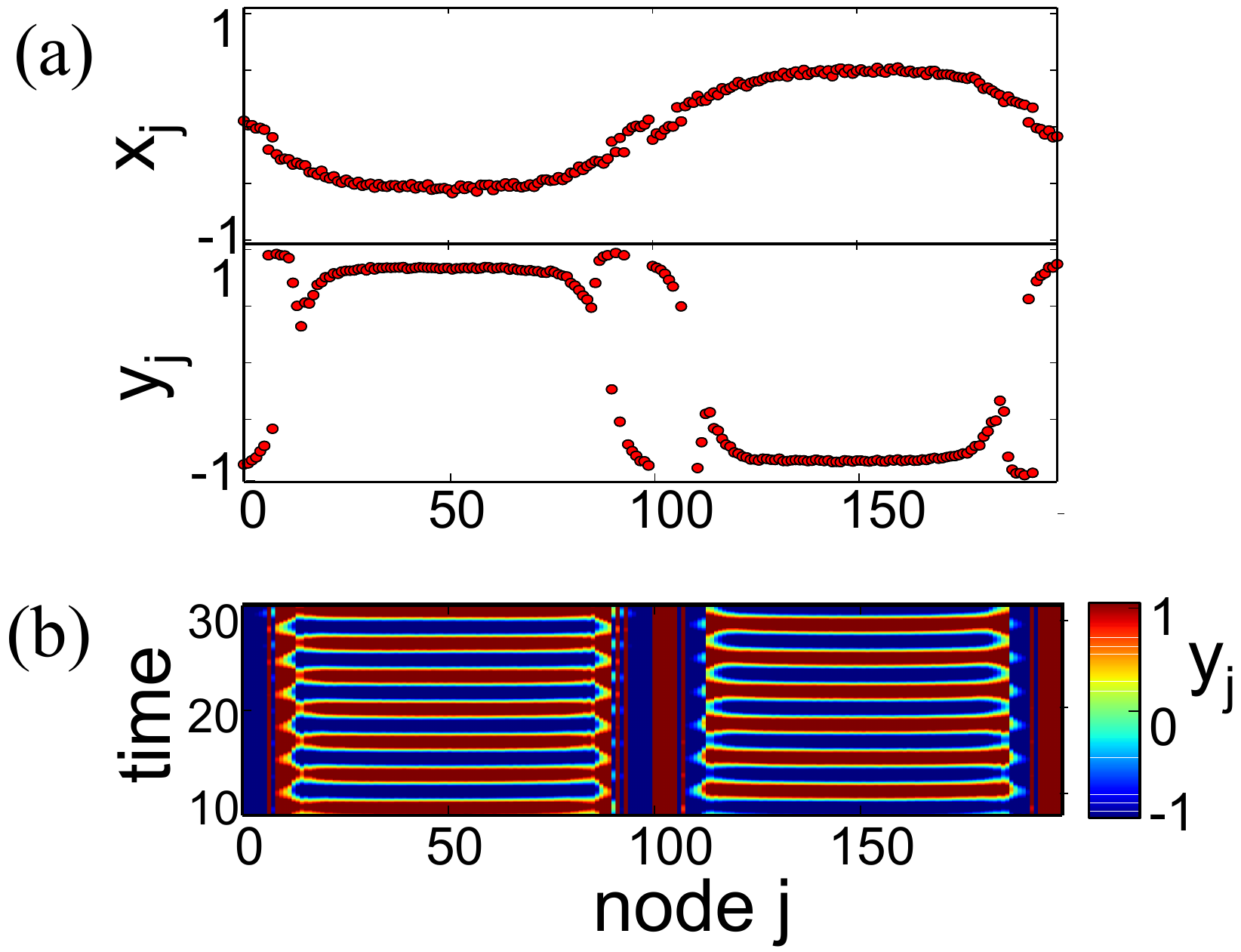}
\caption{(Color online) Amplitude chimera for noise intensity $D\!=\!5\cdot 10^{-3}$:  
(a) snapshot (top: $x_j$, bottom: $y_j$), (b) space-time plot. Parameters: $N\!=\!200$, $P/N\!=\!0.04$, $\sigma\!=\!19$, $\lambda\!=\!1$, $\omega\!=\!2$.
Initial condition: see section \ref{SEC:IC}.
} 
	\label{FIG:AC_De=1}
\end{figure}

In this section we will study the robustness of chimera states with respect to external noise. By using the same initial conditions which lead to amplitude chimera states and chimera death in the deterministic case, we also observe these states in Eq.\,(\ref{EQ:SL_network_det}) in the presence of noise in a wide range of the coupling parameters. The stochastic equations were integrated with the well-established Euler-Maruyama scheme with stepsize $10^{-4}$, see p.\ 340 in \cite{KLO92}.
Figure \ref{FIG:AC_De=1} shows one exemplary configuration for an amplitude chimera which occurs in a system under the impact of noise of intensity $D\!=\!5\cdot 10^{-3}$. \\

\begin{figure}
\centering
\includegraphics[width=0.9\linewidth]{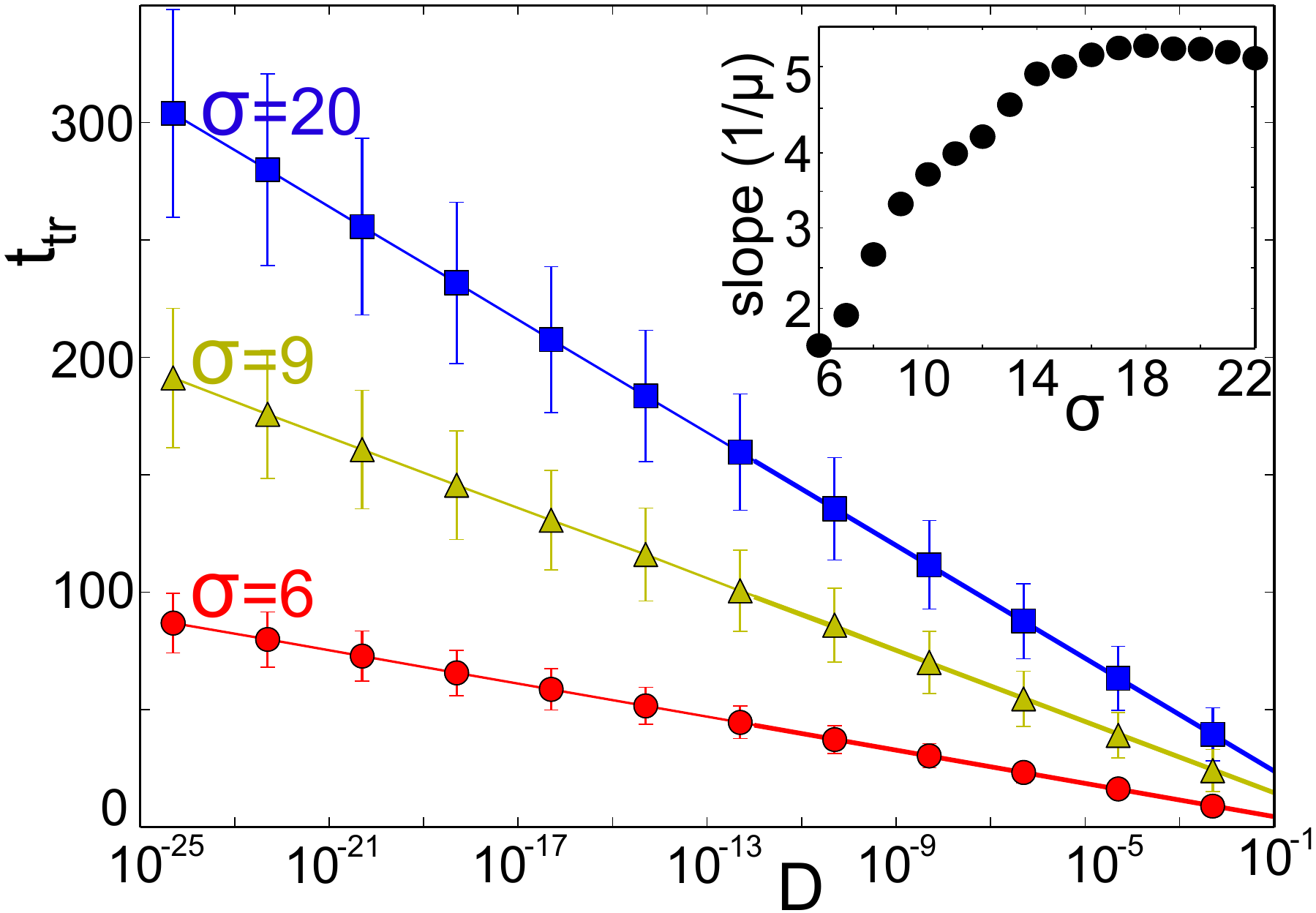}
\caption{(Color online) Transient times of amplitude chimeras $t_{tr}$ vs. noise strength $D$ (log-scaled) for different values of coupling strength $\sigma$. Symbols: average over 50 fully symmetric initial conditions (with $Q=\frac{N}{4}$, each associated with a different realization of the random force $\xi(t)$;
error bars: standard deviations; lines: linear fits from Eq.\,(\ref{EQ:linear-fit}). Inset: slope vs $\sigma$. Parameters: $N\!=\!100$, $P/N\!=\!0.04$, $\lambda\!=\!1$, $\omega\!=\!2$.}
\label{FIG:uncorrelated-noise_trans-D}
\end{figure}

\begin{figure}
\includegraphics[width=\linewidth]{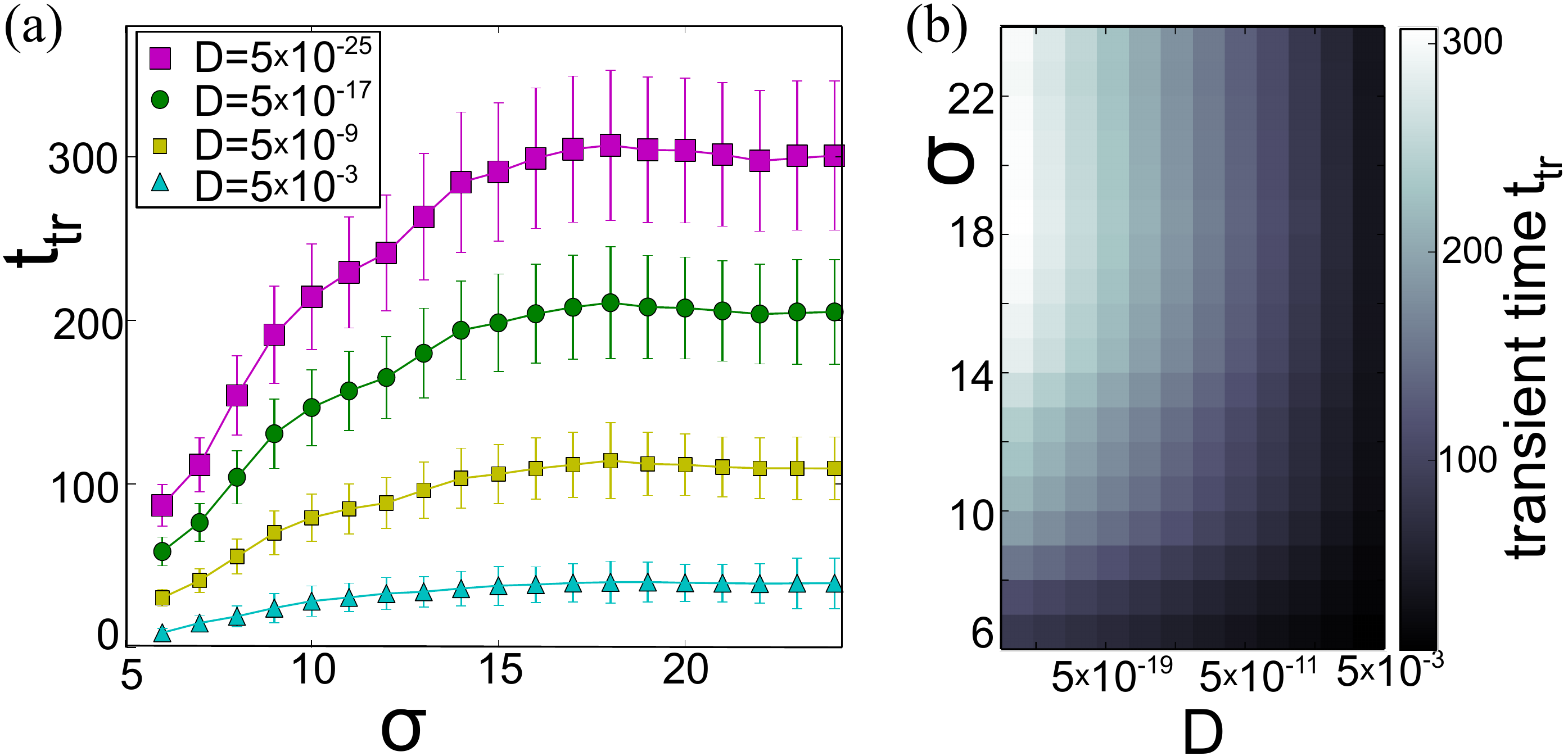}
\caption{(Color online) Transient times of amplitude chimeras $t_{tr}$ averaged over 50 initial conditions and noise realizations (the same set of initial conditions as used in Fig.\,\ref{FIG:uncorrelated-noise_trans-D}): (a) $t_{tr}$ vs. coupling strength $\sigma$ for different noise intensities.
Symbols: mean transient times; error bars: standard deviations. The lines serve as a guide to the eye.
(b) $t_{tr}$ in the plane of coupling strength $\sigma$ and noise intensity $D$. 
Other parameters: $N\!=\!100$, $P/N\!=\!0.04$, $\lambda\!=\!1$, $\omega\!=\!2$.}
\label{transient_vs_sigma-D_imshow}
\label{transient_vs_sigma_sto}
\end{figure}

In general, the transient times of amplitude chimeras decrease with increasing noise intensity. Figure \ref{FIG:uncorrelated-noise_trans-D} shows the average transient times and the corresponding standard deviations in dependence of the noise intensity $D$, for three choices of the coupling strength $\sigma$, in a semi-logarithmic plot. The average is over 50 different fully symmetric initial conditions (with $Q=\frac{N}{4}$, see section \ref{SEC:IC}) drawn from different realizations of the associated random distribution. For each one of those realizations of the initial conditions, a different realization of the Gaussian white noise $\xi_j(t)$ is considered. The average transient times show a clear linear decrease as a function of the logarithmic noise intensity. This behavior is found throughout the range $6 \le \sigma \le 24$, i.e., $t_{tr} = -\frac{1}{\mu} \, ln(D) + \eta$ with slope $-\frac{1}{\mu}$ and axis intercept $\eta$. This gives the scaling law
\begin{equation} \label{EQ:linear-fit}
D  \sim e^{- \mu t_{tr}}.
\end{equation}
The lines in Fig.\,\ref{FIG:uncorrelated-noise_trans-D} show the linear fits, and the inset depicts the slope in dependence on the coupling strength $\sigma$. 
For the same set of 50 initial conditions, Fig.\,\ref{transient_vs_sigma_sto}(a) depicts the mean transient time in dependence of the coupling strength for four different noise intensities $D$, and Fig.\,\ref{transient_vs_sigma-D_imshow}(b) shows a color-coded density plot of the mean transient times of amplitude chimeras in the $(\sigma,D)$-plane. The transient times generally decrease with increasing noise, and increase with increasing coupling strength up to a saturation value at about $\sigma\! \approx \! 15$. The vertical 
error bars in panel (a) show that the transient times are less sensitive to the initial condition, the larger the noise is. We generally find that the spread of the amplitude chimera lifetimes for different initial conditions (and different noise realizations), is smaller with increasing noise strength.\\

\begin{figure}
\centering
\includegraphics[width=0.6\linewidth]{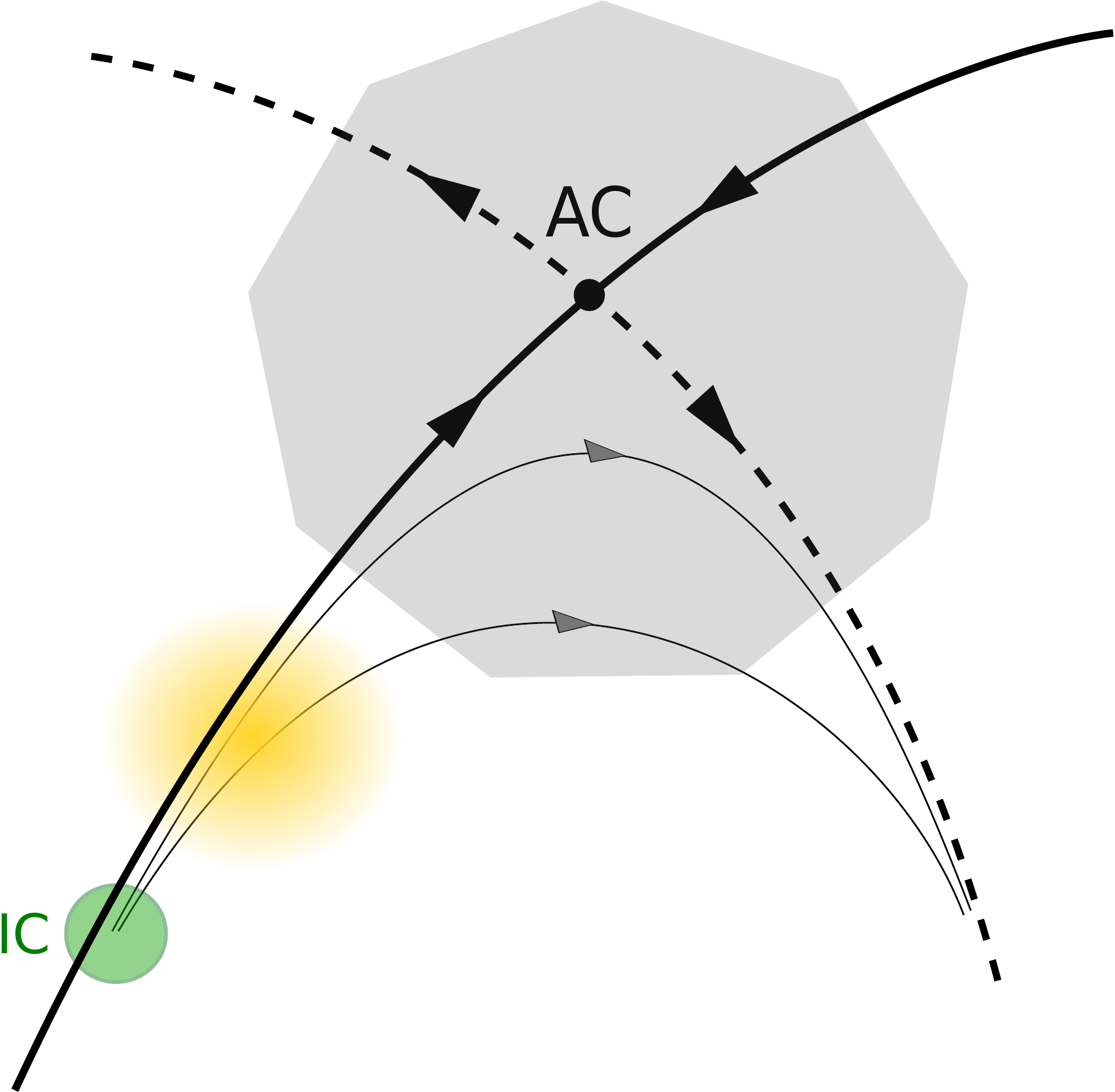}
\caption{(Color online) Schematic phase-space structure of an amplitude chimera (AC) as a saddle-point. Thick solid lines: stable directions, thick dashed lines: unstable directions, thin solid lines: different trajectories, with arrows denoting the direction of time evolution. Grey shaded region: scheme of amplitude chimera configuration, green (dark grey) disk: set of initial conditions (IC), yellow (light grey) area: impact of Gaussian white noise.}
\label{mz}
\label{FIG:AC_Trajectories_sto}
\end{figure}

Transient amplitude chimeras can last for thousands of oscillation periods until they disappear. Even under disturbance by external noise they persist for a significant time. Noise does not essentially change their spatial configuration. 
If noise throws the system onto an adjacent trajectory in the underlying high-dimensional phase space of the network, this does not normally lead to a flow into a completely different direction in phase space. Geometrically speaking, this shows that there are some attracting directions in phase space along which the system dynamics is pushed towards the amplitude chimera. Furthermore, amplitude chimeras can evolve out of initial configurations that do not show the characteristic coexistence of coherent and incoherent domains (see paragraph \ref{IC}). In fact, they can be observed when completely incoherent initial configurations are used, as well as when the initial condition consists of two completely coherent parts. These dynamical properties indicate that the flow within a certain volume of the phase space is directed towards the amplitude chimera state. From the perspective of the amplitude chimera, there must exist some associated ``stable directions''. However, even in the absence of any external perturbation, for all system sizes, the amplitude chimera states disappear after some time, and the system approaches a coherent oscillatory state. Accordingly, there must also exists at least one ``unstable direction'' in phase space. These findings can be explained by the structure of the phase space, which is schematically depicted in Fig.\,\ref{FIG:AC_Trajectories_sto}.\\

The lifetime of amplitude chimeras in the deterministic case strongly depends on the choice of initial conditions as discussed in Sect.
\ref{IC}. In general the sensitivity of chimera states to the initial configurations is explained by the fact that classical chimera states typically coexist with the completely synchronized state, for which the basin of attraction is significantly larger.
For amplitude chimeras, all our numerical results support the idea that amplitude chimera patterns can be seen as a saddle state composed of stable (solid lines in Fig.\,\ref{FIG:AC_Trajectories_sto}) and unstable (dashed lines Fig.\,\ref{FIG:AC_Trajectories_sto}) manifolds. The set of initial conditions leading to amplitude chimeras can be represented as a volume restricted in phase space (green disk in Fig.\,\ref{FIG:AC_Trajectories_sto}). The observed amplitude chimera corresponds to trajectories starting from this set and passing the saddle-point from the stable direction towards the unstable manifold. The lifetime of an amplitude chimera, therefore, depends on the chosen trajectory: the closer to the saddle-point it gets, the longer is the lifetime. In other words, the transient time is determined by the time the system spends in the vicinity of the saddle-point where coherent and incoherent oscillating domains coexist before it escapes to the in-phase synchronized regime along the direction of the unstable manifold. Such a phase space scenario explains the sensitivity of transient times to initial conditions since they determine the particular path the system takes.
With increasing network size $N$, the dimension of the phase space of the network increases, and it becomes more likely that the distance of the initial conditions to the stable manifold is larger, which leads to decreasing transient times for passing the saddle-point. This explains the observed decrease of the lifetimes of the amplitude chimeras with $N$ (Fig.\ref{FIG:mean-transient-times-different-N}). %

Our numerical investigations of the stochastic model Eq.\,(\ref{EQ:SL_network_det}) show that Gaussian white noise dramatically reduces the impact of initial conditions on the lifetime of amplitude chimeras. In more detail, we have tested a set of realizations of initial conditions which lead to significantly different lifetimes of amplitude chimeras without random forcing. In the presence of relatively weak noise $D\!=\!5\cdot 10^{-13}$ all realizations result in amplitude chimeras with similar lifetime. This again supports our view of the amplitude chimera as a saddle-point, and allows for the following explanation. The stochastic force, which continuously perturbs the system, makes it randomly switch between different trajectories close to the saddle-point. Therefore, the system's dynamics is not determined by a single trajectory anymore, but rather affected by a set of trajectories belonging to the $N$-dimensional hyper-sphere. This reduces the sensitivity of the amplitude chimera lifetime to specific initial conditions. In Fig.\,\ref{FIG:AC_Trajectories_sto} the impact of noise is illustrated by yellow shading, denoting the stochastic forces applied to the system at one instant of time.

\section{Maps of dynamic regimes}
\label{SEC:maps}
\begin{figure}
\centering
\includegraphics[width=\linewidth]{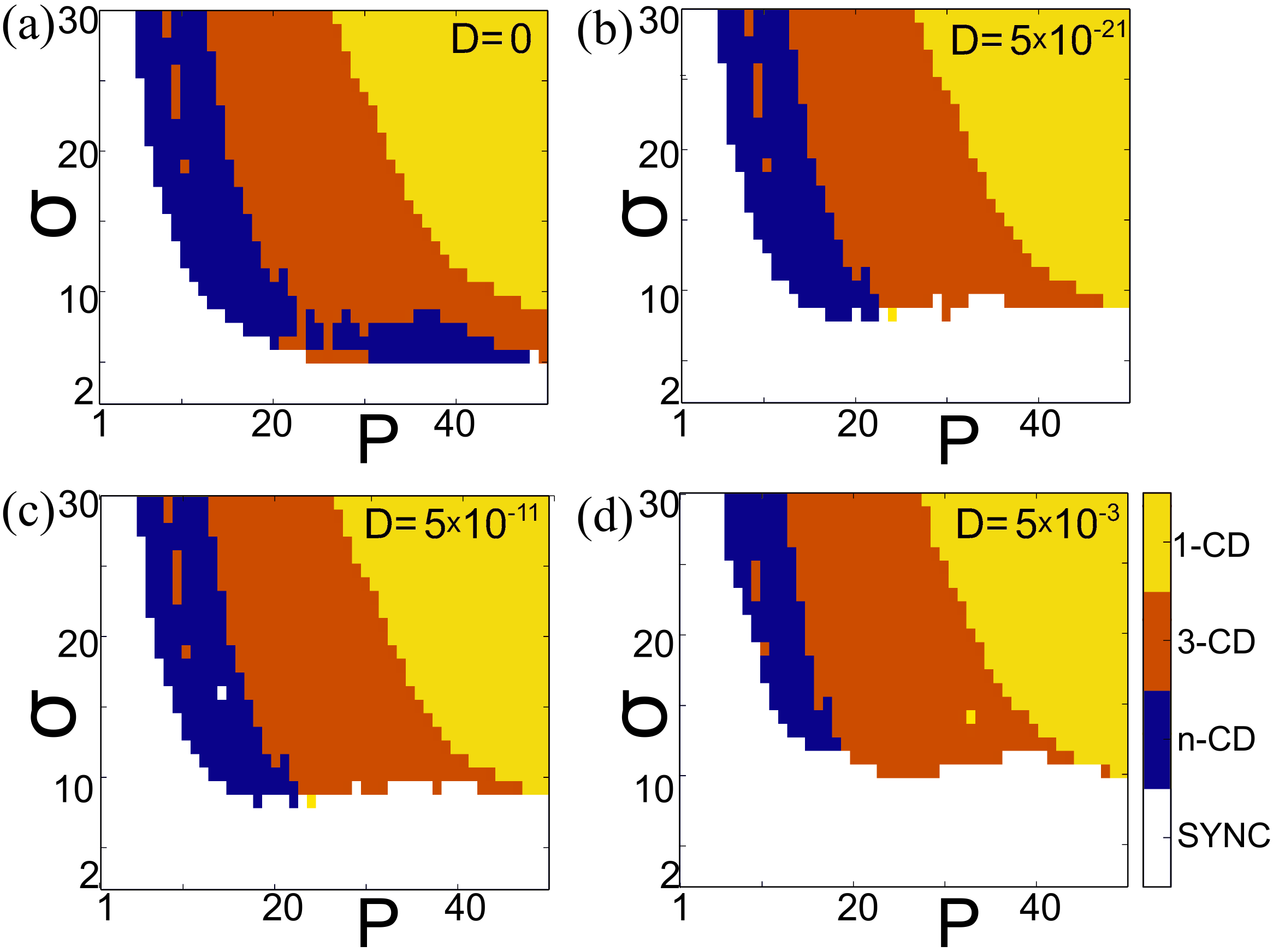}
\caption{(Color online) Map of dynamic regimes in the plane of coupling strength $\sigma$ and coupling range $P$ for noise intensities: (a) $D\!=\! 0$,  (b) $D\!=\! 5\cdot 10^{-21}$, (c) $D\!=\! 5\cdot 10^{-11}$, (d) $D\!=\! 5\cdot 10^{-3}$. Color code: 1-cluster chimera death (1-CD), 3-cluster chimera death (3-CD), multi-cluster chimera death ($n$-CD, $n\!>\!3$), in-phase synchronized oscillations and coherent traveling waves (SYNC). Initial condition: snapshot of an amplitude chimera calculated for $D\!=\!0$, $P\!=\!4$, $\sigma\!=\!14$, $t\!=\!150$. Maximum simulation time: $t\!=\!5000$. Parameters: $N\!=\!100$, $\lambda\!=\!1$, $\omega\!=\!2$.
}
	 \label{FIG:stab-maps_sto}
\end{figure} 
\begin{figure}
\centering
\includegraphics[width=0.9\linewidth]{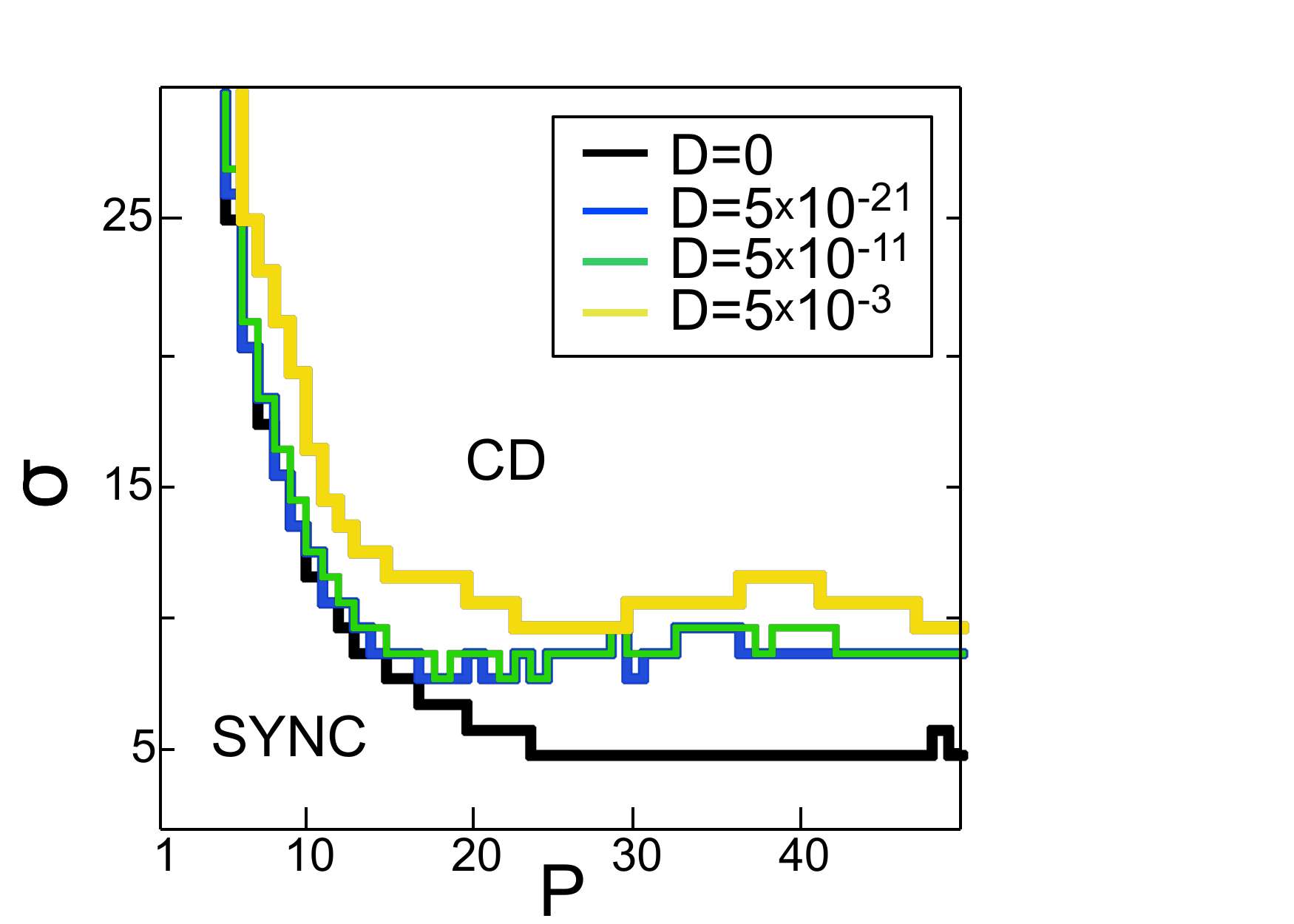}
\caption{(Color online) Boundary between the oscillatory regime and the chimera death regime for different noise intensities $D$,
extracted from the maps of dynamic regimes shown in Fig.\,\ref{FIG:stab-maps_sto}.}
	 \label{FIG:boundaries_stab-maps_sto}
\end{figure} 
For a large range of the coupling parameters $\sigma$ and $P$ we calculate the asymptotically stable state and the transient time of amplitude chimeras for $N\!=\!100$. For each choice of $(\sigma,P)$ we start with the same amplitude chimera configuration as initial condition. For an exemplary initial condition, the results belonging to four different noise intensities are shown in Fig.\,\ref{FIG:stab-maps_sto}.
For very small coupling strength $\sigma$ or very small coupling range $P$ the asymptotic states are coherent oscillatory states, either
in-phase synchronized oscillations or traveling waves (dark blue region, labeled SYNC). For very small coupling range, we observe amplitude chimeras as transients. For larger $\sigma$ and $P$, we find chimera death states (yellow, orange, and blue regions) with one coherent domain (1-CD), or for slightly smaller $P$, with three (3-CD), or more ($n$-CD with $n>3$) coherent domains.
For all noise intensities, there exists a chimera death regime (1-CD, 3-CD, $n$-CD), as well as a coherent oscillatory regime (SYNC).\\
 
The regime of chimera death states is characterized by high multistability, characterized by different initial conditions. The boundary between the oscillatory regime and the chimera death regime is roughly independent of the particular amplitude chimera snapshot used as initial condition. In contrast, for many values of $(\sigma,P)$, the particular type of chimera death depends on the realization of the initial condition. Note that there is nevertheless a clear tendency that the $m$-CD patterns with $(m<k)$ are generally found for larger coupling ranges than the $k$-CD states ($k,m \in \{1,3,n\}$). This tendency is especially pronounced for large coupling strengths.
Noise influences the dynamic regimes in different ways. First, the boundaries between the different cluster types of chimera death appear to be almost unaffected by the applied external noise. We do not observe any noise-induced switching between the different types of chimera death. The applied noise does not influence the asymptotic chimera death state. Second, with increasing noise intensity, the boundary between the oscillatory regime and the oscillation death regime is shifted towards higher coupling strengths. This means that the stochastic force pushes the system out of the deterministic inhomogeneous steady state into the basin of attraction of the stable coherent oscillatory state, and induces oscillations in a parameter regime where in the absence of noise the steady state is a stable asymptotic solution. 
The size of this parameter regime depends on the applied noise intensity. In order to facilitate the comparison, the boundaries between the oscillatory regime and the chimera death regime are depicted for different noise intensities in Fig.\,\ref{FIG:boundaries_stab-maps_sto}. Generally, the stronger the applied noise is, the smaller is the regime of chimera death states.\\

\begin{figure}
\vspace*{0.5cm}
\includegraphics[width=\linewidth]{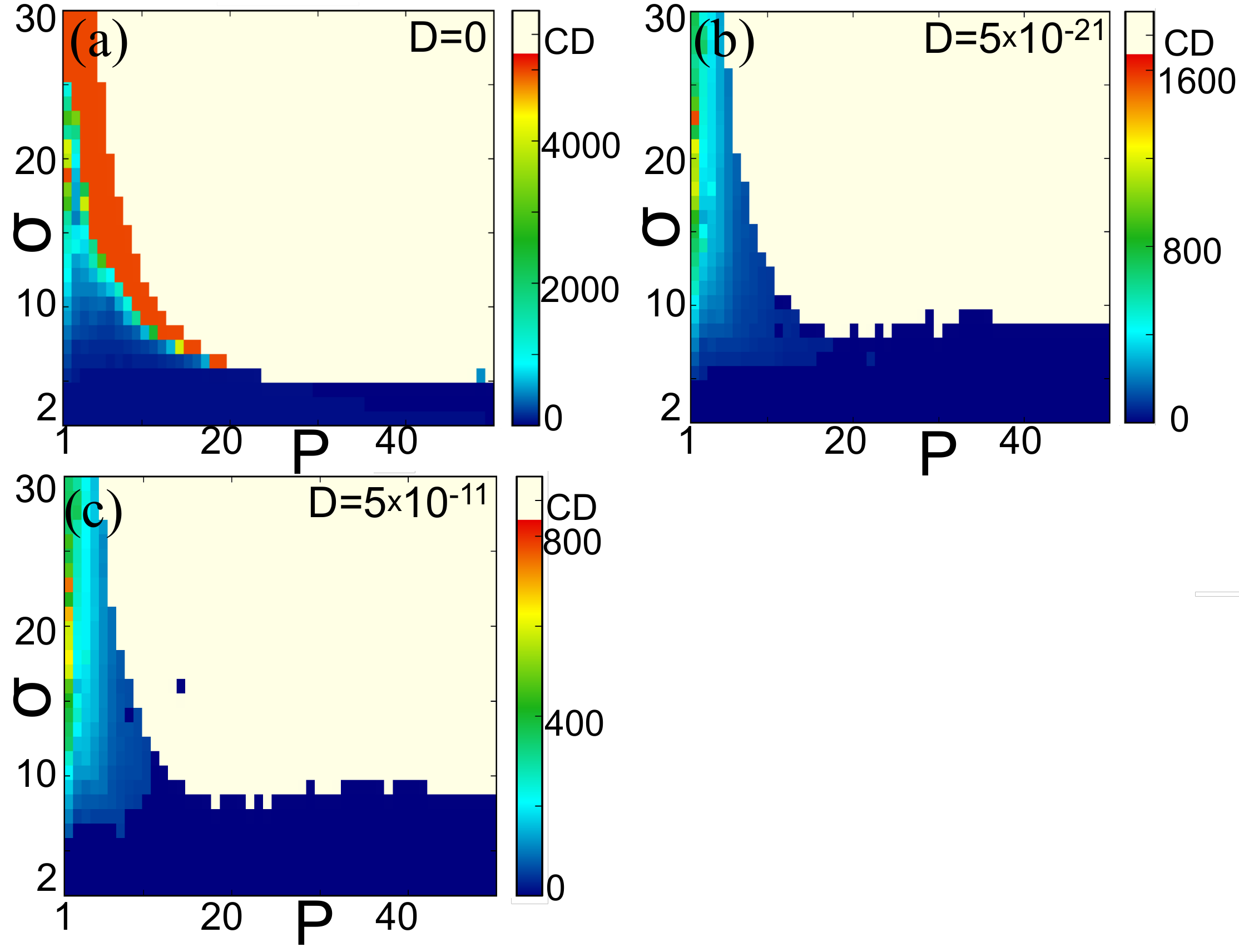}
\caption{(Color online) Transient times of amplitude chimeras $t_{tr}$ in the plane of coupling strength $\sigma$ and coupling range $P$, for the noise intensities: (a) $D\!=\! 0$,  (b) $D\!=\! 5\cdot 10^{-21}$, (c) $D\!=\! 5\cdot 10^{-11}$. System parameters, initial condition and simulation time as in Fig.\,\ref{FIG:stab-maps_sto}.}
\label{FIG:transients_maps}
\end{figure} 
In the oscillatory regime, we observe transient amplitude chimeras. In Fig.\,\ref{FIG:transients_maps} their lifetimes are depicted, obtained from the same simulations described above. One can see that generally the transient time decreases with decreasing coupling strength, and with increasing noise intensity, as shown already in Figs.\,\ref{FIG:uncorrelated-noise_trans-D} and \ref{transient_vs_sigma-D_imshow} for a restricted range of coupling parameters. Note that Fig.\,\ref{FIG:transients_maps}(b) ($D\!=\!5\cdot 10^{-11}$) and (c) ($D\!=\! 5\cdot 10^{-21}$) look very similar up to rescaling of the transient times. This illustrates that the impact of the applied noise upon the dynamics is rather independent of the strength and range of the coupling. \\

In the deterministic case (Fig.\,\ref{FIG:transients_maps}(a)) there is a regime of high values of the coupling strength, at the border between the oscillatory regime and the chimera death regime, where the transient amplitude chimeras last longer than the maximum simulation time of $t\!=\!5000$ (bright orange). For several values of $(\sigma,P)$ in this region, we have simulated much longer time series until $t\!=\!40000$ (more than $12700$ oscillation periods $T$), and have found that the amplitude chimeras persist. However, they disappear much earlier as soon as a tiny amount of external noise is applied. This indicates that the amplitude chimera states are still unstable in this region. The extremely long transient times might simply be related to our choice of initial conditions in the deterministic system. \\

\section{Conclusion}
In this work, the robustness of chimera states with respect to noise has been investigated for a paradigmatic network of oscillators. We have presented numerical results demonstrating that transient amplitude chimeras and chimera death states in a ring network of identical Stuart-Landau oscillators with symmetry-breaking coupling continue to exist in the presence of Gaussian white noise.\\

Transient amplitude chimeras occur in the same range of coupling parameters as in the deterministic case. The key quantity we use to measure their robustness is the transient time. The latter decreases logarithmically with the applied noise intensity. For a constant noise intensity, the transient times increase with the coupling strength up to a saturation value. The width of the incoherent domains relative to the overall system size is not affected by external noise, and is independent of the used initial condition, and the total system size. It increases linearly with the coupling strength as well as with the coupling range.
The amplitude chimeras appear as long lasting transients in systems of all sizes, in contrast to classical phase chimeras whose transient times increase exponentially with the network size \cite{WOL11,ROS14a}. The amplitude chimera lifetime decreases with the network size, but approaches a finite value in the thermodynamic limit.\\

The amplitude chimera lifetimes depend sensitively on the particular realization of the randomized initial condition. We have introduced a class of specially prepared random initial conditions that produce long lasting amplitude chimeras. We have shown that initial configurations that fulfill a symmetry conditions which is also found in oscillation death patterns result in the longest living amplitude chimera transients.\\

We have further introduced of set of global order parameters which are capable to detect the transitions from transient amplitude chimeras to completely in-phase synchronized oscillations or coherent traveling waves. Moreover, they can be used as a measure of the width of the incoherent domains of amplitude chimeras. \\

The chimera death patterns also persist under the impact of stochastic forces. However, the coupling parameter regime where they occur is reduced with increasing noise intensity. The boundary between the coherent oscillatory regime and the chimera death regime is shifted towards higher values of the coupling strength. That means that the system favors oscillatory behavior for a larger coupling parameter regime. In contrast, this boundary appears to be independent on the particular realization of the initial condition. The number of clusters within the coherent domains appears to be unaffected by the external noise, but depends on the particular initial condition.\\

Our numerical findings can be explained by the underlying phase space structure. More specifically, we propose that amplitude chimera states can be represented by saddle states in the phase space of the network. This elucidates the behavior of their lifetime, and explains that generally the initial conditions become less important under the influence of noise.\\

\begin{acknowledgments}
This work was supported by the DFG in the framework of the SFB 910. 
\end{acknowledgments}

\end{document}